%% file: DP2-arXiv.tex
\documentclass[preprint,11pt]{article}
\usepackage{fullpage}
\usepackage{amssymb,amsfonts,amsmath,amscd,dsfont,mathrsfs,bm,amsthm}%,natbib
\usepackage{graphicx,float,psfrag,epsfig,amssymb}
\usepackage[usenames,dvipsnames,svgnames,table]{xcolor}
\definecolor{darkgreen}{rgb}{0.0,0,0.9}
\usepackage[pagebackref,letterpaper=true,colorlinks=true,pdfpagemode=none,citecolor=OliveGreen,linkcolor=BrickRed,urlcolor=BrickRed,pdfstartview=FitH]{hyperref}
\usepackage{hyperref}
\usepackage[tight]{subfigure}
\usepackage{caption}
\let\chapter\section
\usepackage{algorithm}
\usepackage[noend]{algorithmic}
\addtocontents{toc}{\protect\setcounter{tocdepth}{2}}

\input{notation}

\DeclareMathAlphabet{\mathpzc}{OT1}{pzc}{m}{it}
\newtheorem{propo}{Proposition}[section]
\newtheorem{lemma}[propo]{Lemma}
\newtheorem{definition}[propo]{Definition}

\newtheorem{theorem}[propo]{Theorem}

\theoremstyle{definition}

\newtheorem{remark}[propo]{Remark}

\newtheorem{assumption}[propo]{Assumption}

\title{Perishability of Data: Dynamic Pricing\\ under Varying-Coefficient Models}

\author{Adel Javanmard\\
 Department of Data Sciences and Operations\\
 Marshall School of Business\\
 University of Southern California, Los Angeles, CA 90089\\ 
 {\small{ajavanma@usc.edu}}
            }

\begin{document}

\maketitle

%\begin{abstract}
\begin{abstract}
We consider a firm that sells a large number of products to its customers in an online fashion. Each product is described by a high dimensional feature vector, and the market value of a product is assumed to be linear in the values of its features. Parameters of the valuation model are unknown and can change over time.  The firm sequentially observes a product's features and can use the historical sales data (binary sale/no sale feedbacks) to set the price of current product, with the objective of maximizing the collected revenue. We measure the performance of a dynamic pricing policy via regret, which is the expected revenue loss compared to a clairvoyant that knows the sequence of model parameters in advance. 

We propose a pricing policy based on projected stochastic gradient descent (PSGD) and characterize its regret in terms of time $T$, features dimension $d$, and the temporal variability in the model parameters, $\delta_t$. 
We consider two settings. In the first one, feature vectors are chosen antagonistically by nature and we prove that the regret of PSGD pricing policy is of order $O(\sqrt{T} + \sum_{t=1}^T \sqrt{t}\delta_t)$. In the second setting (referred to as stochastic features model), the feature vectors are drawn independently from an unknown distribution. We show that in this case, the regret of PSGD pricing policy is of order $O(d^2 \log T + \sum_{t=1}^T t\delta_t/d)$.

\end{abstract}

%\KEYWORDS{Revenue Management, Dynamic Pricing, High-dimensional Regression, Maximum Likelihood,  Sparsity, Hypothesis Testing}

\section{Introduction} \label{sec:intro}
Motivated by the prevalence of online marketplaces, we consider the problem of a firm selling a large number of products, that are significantly differentiated from each other, to customers that arrive over time. The firm needs to price the products in a dynamic manner, with the objective of maximizing the expected revenue. 

The majority of work in dynamic pricing assume that a retailer sells \emph{identical} items to its customers~\cite{BesbesZ09,farias2010dynamic,broder2012dynamic,den2013simultaneously,WangDY14}. Recently, feature-based models have been used to model the products differentiation by assuming that each product is described by vectors of high-dimensional features. These models are suitable for business settings where there are an enormous number of distinct products. One important example is online ad markets. In this context, products are the impressions (user view) that are sold by the web publisher to advertisers.
Due to the ever-growing amount of data that is available on the Internet, for each impression there is large number of associated features, including demographic information, browsing history of the user, and context of the webpage. Many other online markets, such as Airbnb, eBay and Etsy also have a similar setting in which products to be sold are highly differentiated. For example, in the case of Aribnb, the products are ``stays" and each is characterized by a large number of features including space properties, location, amenities, house rules, as well as arrival dates, events in the area, availability of near-by hotels, etc~\cite{Airbnb}. 

Here, we consider a feature-based model that postulates a linear relation between the market value of each product and its feature values. Further, from the firm's perspective, we treat distinct buyers independently, and hereafter focus on a single buyer.  Put it formally, we start with the following model for the buyer's valuation:
\begin{align}\label{eq:model1}
v(x_t) = \<x_t, \th\> + z_t\,,
\end{align}
where $x_t\in \reals^d$ denotes the product feature vector, $\th$ represents the buyer's preferences and $z_t$, $t\ge 1$ are idiosyncratic shocks, referred to as noise, which are drawn independently and identically from a zero mean distribution. For two vectors $a,b$, we write $\<a,b\>$ to refer to their inner product. Feature vectors $x_t$ are observable, while model parameter $\th$ is a-priori unknown to the firm (seller). Therefore, the buyer's valuation $v(x_t)$ is also hidden from the firm.
  
Parameters of the above model represents how different features are weighted by the buyer in assessing the product. Considering such model, a firm can use historical sales data to estimate parameters of the valuation model, while concurrently collecting revenue from new sales. In practice, though, the buyer's valuation of a product will change over time and this raises the concern of \emph{perishability} of sales data.  

In order to capture this point, we consider a richer model with varying coefficients:
\begin{align}\label{eq:VCmodel}
v_t(x_t) = \<x_t, \th_t\> + z_t\,.
\end{align}
Model parameters $\theta_t$ may change over time and as a result, valuation of a product depends on both the product feature vector and the time index.  

We study a dynamic pricing problem, where at each time period $t$, the firm has a product to sell and after observing the product feature vector $x_t$, posts a price $p_t$. If the buyer's valuation is above the posted price, $v_t(x_t) \ge p_t$, a sale occurs and the firm collects a revenue of $p_t$.
If the posted price exceeds the buyer's valuation, $p_t > v_t(x_t)$, no sale occurs. Note that at each step, the firm has access to the previous feedbacks (sale/no sale) from the buyer and can use this information in setting the current price.

In this paper, we will analyze the varying-coefficient model~\eqref{eq:VCmodel} and answer two fundamental questions:
\begin{quote}
First, what is the value of knowing the sequence of model parameters $\theta_t$; in other words, what is the expected revenue loss (regret) compared to the clairvoyant policy that knows the parameters of the valuation model in advance? Second, what is a good pricing policy? 
\end{quote}

The answer to the first question intrinsically depends on the temporal variability in the sequence $\th_t$. If this variation is very large, then there is not much that can be learnt from previous feedback on the buyer's behavior and the problem turns into a random price experimentation. On the other hand, if all of the parameters $\th_t$ are the same, then this feedback information can be used to learn the model parameters, which in turn helps in setting the future prices. In this case, an algorithm that performs a good balance between price exploration and best-guess pricing (exploitation) can lead to a small regret. In this work, we study this trade-off through a projected stochastic gradient descent algorithm and investigate the effect of variations of the sequence of $\th_t$ on the regret bounds. 

Feature-based models have recently attracted interest in dynamic pricing. \cite{amin2014repeated} studied a similar model to~\eqref{eq:model1} (without the noise terms $z_t$), where the features $x_t$ are drawn from an unknown i.i.d distribution. A pricing strategy was proposed based on stochastic gradient descent, which results in a regret of the form $O(T^{2/3} \sqrt{\log T})$. This work also studied the problem of dynamic incentive compatibility in repeated posted-price auctions. Subsequently,~\cite{cohen2016feature} studied model~\eqref{eq:model1}, wherein the feature vectors $x_t$ are chosen antagonistically by nature and not sampled i.i.d. This work proposes a pricing policy based on the ellipsoid method from convex optimization~\cite{boyd2004convex} with a regret bound of $O(d^2 \log(T/d))$, under a low-noise setting. More accurately, the regret scales as $O(d^2 \log(\min\{T /d, 1/\delta\}) + d\delta T)$, where $\delta$ measures the noise magnitude: in case of  bounded noise, $\delta$ represents the uniform bound on noise and in case of gaussian noise with variance $\sigma^2$, it is defined as $\delta = 2\sigma \sqrt{\log(T)}$. In~\cite{lobel2016multidimensional}, the regret bound of this policy was improved to $O(d \log T)$, under the noiseless setting.  
In~\cite{JavanmardNazerzadeh}, authors study and highlight the role of the structure of demand curve in dynamic pricing. They introduce model~\eqref{eq:model1}, and assume that the feature vectors $x_t$ are drawn i.i.d. from an unknown distribution. Further, motivated by real-world applications, it is assumed that the parameter vector $\th$ is sparse in the sense that only a few of its entries are nonzero. A regularized log-likelihood approach is taken to get an improved regret bound of order $s_0 (\log (d)+\log(T))$. We add to this body of work by considering feature-based models for valuation of products whose parameters vary over time. 

Time-varying demand environments have also been studied recently by~\cite{Keskin-TVC}. Explicitly, they consider a firm that sells one type of product to customers that arrive over a time horizon. After setting price $p_t$, the firm observes demand $D_t$ given by $D_t = \alpha_t + \beta_t p_t+\epsilon_t$, where $\alpha_t, \beta_t \in \reals$ are the unknown parameters of the demand model and $\epsilon_t$ are the unobserved demand shocks (noise). By contrast, in this work we consider different products, each characterized by a high-dimensional feature vector. Further, the seller only receives a binary feedback (sale/no sale) of the customer's behavior at each step, rather than observing the customer's valuation.

\subsection{Organization of paper and our main contributions}
The remainder of this paper is structured as follows. In Section~\ref{sec:model}, we formally define the model and formulate the problem. Technical assumptions and the notion of regret will be discussed in this section. We next propose a pricing policy based on projected stochastic gradient descent (PSGD) applied to the log-likelihood function. At each time period $t$, it returns an estimate $\hth_t$. The price $p_t$ is then set to the optimal price as if $\hth_t$ was the actual parameter $\th_t$.  We next analyze the regret of our PSGD algorithm. Let $\delta_t = \|\th_{t+1}-\th_t\|$ be the variation in model parameters at time period $t$.  In Section~\ref{sec:regret}, we consider the setting where the product feature vectors $x_t$ are chosen antagonistically by nature and show that the regret of PSGD algorithm is of order $O(\sqrt{T} + \sum_{t=1}^T \sqrt{t} \delta_t)$. Interestingly, this bound is independent of the dimension $d$, which is a desirable property of our policy for high-dimensional applications. We next, in Section~\ref{stochastic}, consider a stochastic features model, where the feature vectors $x_t$ are drawn independently from an unknown distribution (cf. Assumption~\ref{SMF}). Under this setting, we show that the regret of PSGD is of order $O(d^2 \log T + \sum_{t=1}^T t\delta_t/d)$. Note that setting $\delta_t=0$ corresponds to model~\eqref{eq:model1} and our PSGD pricing obtains a logarithmic regret in $T$. Section~\ref{sec:thm} is devoted to the proof of main theorems and the main lemmas are proved in Section~\ref{sec:lem}. Finally, proof of several technical steps are deferred to Appendices.

\subsection{Related literature}
Our works is at the intersection of dynamic pricing, online optimization and high-dimensional statistics. In the following, we briefly discuss the work most related to ours from these contexts.
\smallskip

\noindent{\bf Feature-based dynamic pricing.} 
Recent papers on dynamic pricing consider models with features/covariates, motivated in part by new advances in big data technology that allow firms to collect large amount of fine-grained information. In the introduction, we discussed the work~\cite{amin2014repeated,JavanmardNazerzadeh,cohen2016feature} which are closely related to our setting. Another recent work on feature-based dynamic pricing is~\cite{qiang2016dynamic}. In this work, authors consider a model where the seller observes the demand entirely, rather than a binary feedback as in our setting. A greedy iterative least squares (GILS) algorithm is proposed that at each time period estimates the demand as a linear 
function of price by applying least squares to the set of prior
prices and realized demands. The work underscores the role of feature-based approaches and show that they create enough price dispersion to achieve a regret of $O(\log (T))$. This is closely related to the work of \cite{den2013simultaneously} and \cite{keskin2014dynamic}  in dynamic pricing (without  demand covariates) that demonstrate the GILS is suboptimal and propose methods to integrate forced price-dispersion with GILS to achieve optimal regret.
\smallskip

\noindent{\bf Online optimization.}
This field offers a variety of tools for sequential prediction, where an agent measures its predictive performance according to a series of convex functions. Specifically, there is a sequence of a priori unknown 
reward functions $f_1, f_2, f_3, \dotsc$ and an agent must make a sequence of decisions: at each time period $t$, he selects a point $z_t$ and a loss $f_t(z_t)$ is incurred. Note that the function $f_t$ is not known to agent at step $t$, but he has access to all previous functions $f_1, \dotsc, f_{t-1}$. First order methods, like online gradient descent (OGD) or online mirror descent (OMD) only use the gradient of previous function at the selected points, i.e., $\partial f_t(z_t)$. The notion of regret here is defined by comparing the agent with the best fixed comparator~\cite{shalev2011online}.

\cite{Hall-DGD} proposed dynamic mirror descent that is capable of adapting adapts to a possibly non-stationary environment. In contrast to OMD~\cite{beck2003mirror,shalev2011online}, the notion of regret is defined more generally with respect to the best comparator ``sequence". 

It is worth noting that the general framework of online learning does not directly apply to our problem. To see this, we define the the loss $f_t$ to be the negative of the revenue obtained in time period $t$, i.e., $f_t = -p_t \ind(p_t\ge v_t)$. Then 1) the loss functions are not convex; 2) the (first order information) of previous loss functions depend on the corresponding valuations $v_1, \dotsc, v_{t-1}$ which are never revealed to the seller.
That said,  we borrow some of the techniques from online optimization in proving our results. (See proof of Lemma~\ref{lem:PE}.)
   
\smallskip

\noindent{\bf High-dimensional statistics.}  Among the work in this area, perhaps the most related one to our setting is the problem of 1-bit compressed sensing~\cite{plan2013one, plan2013robust,ai2014one,bhaskar20151}. In this problem, a set of linear measurements
are taken from an unknown vector and the goal is to recover this vector having access to the sign of these measurements (1-bit information). This is related to the dynamic pricing problem on model~\eqref{eq:model1}, as the seller observes 1-bit feedback (sale/no sale from previous time periods). However, there are a few important differences between these two problem that are worth noting: 1) In dynamic pricing, the crux of the matter is  the decisions (prices) made by the firm. Of course this task entails learning the model parameters and therefore the firm gets into the realm of exploration (learning) and exploitation (earning revenue). By contrast, 1-bit compressed sensing is only a learning task; 2)  In dynamic pricing, the prices are set based on the previous (sale/no sale) feedbacks. Therefore, the feedbacks are inherently correlated  and this makes the learning task challenging. However, in 1-bit compressed sensing it is assumed that the measurements (and therefore the observed signs ) are independent; 3) The majority of work on 1-bit compressed sensing consider an offline setting, while in the dynamic pricing, decision are made in an online manner.

\section{Model} \label{sec:model}

We consider a pricing problem faced by a firm that sells products in a sequential manner. At each time period $t=1,2,\cdots,T$ the firm has a product to sell and the product is represented by an {\em observable} vector of features (covariates) $x_t \in \cX \subseteq \reals^d$.  The length of the time horizon, denoted by $T$, is \emph{unknown} the to the firm and the set $\cX$ is bounded.

The product at time $t$ has a market value $v_t = v_t (x_t)$, depending on both $t$ and $x_t$, which is {\em unobservable}.
At each period $t$, the firm (seller) posts a price $p_t$. If $p_t\le v_t$, a sale occurs, and the firm collects revenue $p_t$. If the price is set higher than the market value, $p_t>v_t$, no sale occurs and no revenue is generated. 
The goal of the firm is to design a pricing policy that maximizes the collected revenue.

We assume that the market value of a product is a linear function of its covariates, namely 
\begin{align}\label{eq:model}
v_t(x_t) = \<\th_t, x_t\>+z_t\,.
\end{align}
Here, $\theta_t$ and $x_t$ are $d$-dimensional and $\{z_t\}_{t\ge 1}$ are idiosyncratic  shocks, referred to as noise, which are drawn independently and identically from a zero-mean distribution over $\reals$. We denote its cumulative distribution function by $F$, and the corresponding density by $f(x) = F'(x)$. Note that the noise can account for the features that are not measured. We refer to~\cite{keskin2014dynamic,debBoerZwart2014,qiang2016dynamic} for a similar notion of demand shocks.

The sequence of parameters $\bth= (\th_1,\th_2,\dotsc)$ is \emph{unknown} to the firm and they may vary across time. This paper focuses on arbitrary sequences $\bth$ and propose an efficient algorithm whose regret scale gracefully in time and the temporal variability in the sequences of $\th_t$. The regret is measured with respect to the clairvoyant policy that knows the sequence $\bth$ in advance.
We will formally define the regret in Section~\ref{sec:Benchmark}.

We let $y_t$ be the response variable that indicates whether a sale has occurred at period $t$:
\begin{align}
y_t = \begin{cases}
+1&\text{ if }v_t \ge p_t\,,\\
-1&\text{ if }v_t <p_t\,.
\end{cases}
\end{align}
Note that the above model for $y_t$ can be represented as the following probabilistic model:
\begin{align}\label{eq:prob-model}
y_t = \begin{cases}
+1&\text{ with probability }\, 1- F\left(p_t-\<\th_t,x_t\>\right)\,,\\
-1&\text{ with probability }\, F\left(p_t- \<\th_t, x_t\>\right)
\end{cases}
\end{align}
%

%Our proposed algorithm exploits the structure (sparsity) of the feature space to improve its performance. % namely, if only a few of the covariates are predictor of the market value. 
%To this aim, let $s_0$ denote the number of nonzero coordinates of $\tth$, i.e., $s_0  =  \|\tth\|_0 = \sum_{j=1}^d \ind(\tth_j\neq 0)$. We remark  that $s_0$ is a-priori unknown to the seller.  

%Namely, In particular, we suppose that only a few of the covariates are predictor of the market value of a product and ergo the parameter vector $\tth$ is sparse. Throughout, we let $s_0$ be the number of nonzero coordinates
%of $\tth$, $s_0  = \|\tth\|_0$.  

\subsection{Technical assumptions and notations}
For a vector $v$, we write $\|v\|_p$ for the standard $\ell_p$ norm of a vector $v$, i.e., $\|v\|_p = (\sum_i |v_i|^p)^{1/p}$.
Whenever the subscript $p$ is not mentioned it is deemed as the $\ell_2$ norm. For a matrix $A$, $\|A\|$ denotes its $\ell_2$ operator norm.
For two vectors $a, b$, we use the notation $\<a,b\>$ to refer to their inner product.

To simplify the presentation, we assume that $\|x_t\| \le 1$, for all $x_t\in \cX$, and  $\|\th_t\|\le \l1u$ for a known constant $\l1u$. 
We denote by $\Theta$ the $d$-dimensional $\ell_2$ ball of radius $\l1u$ (In fact, we can take $\Theta$ to be any convex set that contains 
parameters $\th_t$. The size of $\Theta$ effects our regret bounds up to a constant factor.)

We also make the following assumption on the distribution of noise $F$.
\begin{assumption}\label{ass1}
The function $F(v)$ is strictly increasing. Further, $F(v)$ and $1-F(v)$ are log-concave in $v$.
\end{assumption}

Log-concavity is a widely-used assumption in the economics literature~\cite{bagnoli2005log}.
Note that if the density $f$ is symmetric and the distribution $F$ is log-concave, then $1-F$ is also log-concave.
Assumption~\ref{ass1} is satisfied by several common probability distributions including normal, uniform, Laplace, exponential, and logistic.
Note that the cumulative distribution function of all log-concave densities is also log-concave~\cite{boyd2004convex}.

We use the standard big-$O$ notation. In particular $f(n) = O(g(n))$
if there exists a constant $C > 0$ such that $|f(n)| \le Cg(n)$ for all $n$ large enough. We also use $\reals_{\ge 0}$ to refer to the set of non-negative real-valued numbers.

%\begin{assumption}\label{ass2}
%Assume that the distribution of covariates, $\pX$ has a bounded support $\cX$. Let $\Sigma$ denote the covariance matrix of distribution $\pX$. We assume that there exist constants  $C_{\min}$ and $C_{\max}$ such that for every eigenvalue $\sigma$ of $\Sigma$, we have  $0 < C_{\min}\le \sigma < C_{\max} < \infty$.
%\end{assumption}

%A generic example of a probability distribution with positive definite covariance is the case that $\pX$ is bounded below from zero on an open set around the origin. In particular, $\pX \ge a$ for all $\|x\|_\infty \le c$ for some constants $a,c>0$~\cite{bastani2016decision}. This condition holds for many common probability distributions, such as uniform, truncated normal, and in general truncated version of many more distributions.

\subsection{Benchmark policy and regret minimization}\label{sec:Benchmark}
For a pricing policy, we measures its performance via the notion of regret, which is the expected revenue loss compared to an oracle that knows the sequence of model parameters in advance (but not the realizations of $\{z_t\}_{t\ge 1}$).We first characterize this benchmark policy.

Using Eq.~\eqref{eq:model}, the expected revenue from a posted price $p$ is equal to $p\times\prob(v_t\ge p)= p(1-F(p-\th_t\cdot x_t))$. First order condition for the optimal price $p^*(x_t,\th_t)$ reads
\begin{align}\label{eq:opt-p}
\popt(x_t,\th_t) = \frac{1-F\left(\popt(x_t,\th_t)-\<\th_t, x_t\>\right)}{f\left(\popt(x_t,\th_t)-\<\th_t, x_t\>\right)}\,.
\end{align}
To lighten the notation, we drop the arguments $x_t$, $\th_t$ and denote by $p^*_t$ the optimal price at time $t$.

We next recall the \emph{virtual valuation} function, commonly used in mechanism design~\cite{Myerson81}:
$$\varphi(v)\equiv v - \frac{1-F(v)}{f(v)}\,.$$ 
Writing Eq.~\eqref{eq:opt-p} in terms of function $\varphi$, we get
$$\<\th_t, x_t\> + \varphi\left(\popt_t-\<\th_t, x_t\>\right)=0\,.$$
In order to solve for $p^*_t$, we define the pricing function $g$ as follows:
%\begin{align} \label{eq:phi}
%\varphi(v) \equiv v - \frac{1-F(v)}{f(v)}
%\end{align} 
%
%Note that function $\varphi$ corresponds to the \emph{virtual valuation} function commonly used in mechanism design~\cite{Myerson81}.
%By Assumption~\ref{ass1}, $\varphi$ is injective and hence we can define function $g$ as follows
%
\begin{align}\label{eq:g}
g(v) \equiv v + \varphi^{-1}(-v)\,.
\end{align}
By Assumption~\ref{ass1}, $\varphi$ is injective and hence $g$ is well-defined.
Further, it is easy to verify that $g$ is non-negative. 
Using the definition of $g$ and rearranging the terms we obtain
\begin{align}\label{eq:popt}
\popt_t = g(\<\th_t, x_t\>)\,.
\end{align}

%Let $B_q(\l1u)$ denote the $\ell_q$-ball of unit \hn{not unit?} radius. 
%%
%\begin{align}
%B_q(\l1u) = \{\th\in \reals^d:\quad \|\th\|_q^q \hn{only subscript?} = \sum_{j=1}^d |\beta_j|^q \le \l1u\}\,.
%\end{align}
%%
%In the limiting case $q=0$, we define the $\ell_0$-ball
%%
%\begin{align}
%B_0(s) = \{\th\in \reals^d:\quad \sum_{j=1}^d \ind(\beta_j\neq 0) \le s\}\,,
%\end{align}
%%
%which corresponds to the set of vectors $\th$ with at most $s$ non-zero coordinates. Define set $\Omega = B_0(s_0) \cap B_1(\l1u)$, where the dependence on $s_0$ and $\l1u$ is implicit in the notation.

The performance metric we use in this paper is the worst-case regret with respect to a clairvoyant policy that knows the sequence $\bth$ in advance. Formally, for a policy $\pi$ to be the seller's policy that sets price $p_t$ at period $t$, the worst-case regret is defined over $T$ periods is defined as:
\begin{align}\label{eq:Regret_def}
\Reg^\pi(T) \equiv \sup\, \left\{\Delta^{\pi}_{\bth,\bx}:\, \th_t\in \Theta, \, x_t \in \cX \right\}\,,
\end{align}
where for $T\ge 1$, $\bth= (\th_1,\dotsc, \th_T)$ and ${\bx} = (x_1, x_2, \dotsc, x_T)$,
\begin{align}
 \Delta^\pi_{\bth,\bx}(T) = \E_{\bth,\bx} \left[\sum_{t=1}^T \bigg(p^*_t \ind(v_t \ge p^*_t) - p_t \ind(v_t \ge p_t) \bigg)\right]\,.
\end{align}
Here the expectation $\E_{\bth,\bx}$ is with respect to the distributions of idiosyncratic noise, $z_t$. Note that $v_t$, $p_t$, and $p^*_t$ depend on $\bth$ and $\bx$.

\section{Pricing policy} \label{sec:pricing_alg}

\begin{algorithm}[t]
\caption*{{\bf PSGD (Projected stochastic gradient descent) pricing policy}}\label{alg-linear}
\begin{algorithmic}[1]

\REQUIRE{\bf (at time $0$)} function $g$, set $\Theta$,\hspace{2cm}
\REQUIRE{\bf (arrives over time)} covariate vectors $\{x_t\}_{t\in \naturals}$ 

\ENSURE prices $\{p_t\}_{t\in \naturals}$ 

%\hnx{Is is possible to drop the line numbers?}
\STATE $p_1 \leftarrow 0$ and initialize $\hth_1 \in \Theta$

\FOR{$t = 1,2,3,\dots$}

\STATE Set $\hth_{t+1}$ according to the following rule:
\begin{eqnarray}\label{eq:ML}
\hth_{t+1} = \Pi_\Theta(\hth_t - \eta_t \nabla\ell_t(\hth_t)) 
\end{eqnarray}
with 
\begin{align} \label{eq:lt}
\ell_t(\th) = - \ind(y_t =1) \log (1-F(p_t - \<x_t, \th\> )) - \ind(y_t =-1) \log (F(p_t - \<x_t, \th\> )) 
\end{align}
%
%\begin{align} \label{eq:log_likelihood}
%\cL(\th) = - \frac{1}{\tau_{k-1}}\sum_{t\in \cT_{k-1}} \bigg\{\ind(y_t =1) \log (1-F(p_t - \th\cdot x_t )) + \ind(y_t =-1) \log (F(p_t - \th\cdot x_t )) \bigg\}\,.
%\end{align}
%

\STATE Set price $p_{t+1}$ as
\begin{align}\label{eq:price}
p_{t+1} \leftarrow g(\<x_{t+1}, \hth_{t+1}\>) 
\end{align}

\ENDFOR
\end{algorithmic}
\end{algorithm}

% {\bf Regularized Maximum Likelihood (RML)}.

Our dynamic pricing policy consists of a projected gradient descent algorithm to predict parameters $\hth_t$. With each new product, it computes
the negative gradient of the loss and shirts its prediction in that direction. The result is projected onto set $\Theta$ to produce the next prediction.
The policy then sets the prices as $p_t = g(\<x_t,\hth_t\>)$. Note that by Eq.~\eqref{eq:g}, $p_t$ is the optimal price if $\hth_t$ was the true parameter $\th_t$. 
Also, by log-concavity assumption on $F$ and $1-F$, the function $\ell_t(\th)$ is convex.

In projected gradient descent, the sequence of step sizes $\{\eta_t\}_{t\ge 1}$ is an arbitrary sequence of non-increasing values. 
In Sections~\ref{sec:regret} and \ref{stochastic}, we analyze the regret of our pricing policy and provide guidelines for choosing step sizes.

%We restrict our attention to $F$ such that the above quantities are well defined. Specifically, we assume that $F$ is nonzero in $[-M,M]$.\hn{This should follow from %Assumption~\ref{ass1}?}

%The quantities $u_M$ and $\ell_M$ appear in our choice of regularization parameters $\lambda$ and also the error bound for estimating $\tth$.  

\subsection{Regret analysis}\label{sec:regret}
We first define a few useful quantities that appear in our regret bounds. 
Define
\begin{eqnarray} 
M &\equiv& \l1u + \varphi^{-1}(0)\,,\label{eq:M}\\
u_{M} &\equiv& \sup_{|x|\le M} \left\{\max\Big\{ - \dx \log F(x) , -\dx \log(1-F(x)) \Big\}\right\}\,,\label{eq:uM}\\
\ell_{M} &\equiv& \inf_{|x|\le M} \left\{\min\Big\{ -\ddx \log F(x) , -\ddx\log(1-F(x)) \Big\}\right\}\,,\label{eq:lM}
\end{eqnarray}
where the derivatives are with respect to $x$.
We note that $M$ is an upper-bound on the maximum price offered and also, by the log-concavity property of $F$ and $1-F$, we have 
$$u_M = \max\Big\{-\dx\log F(-M) , -\dx\log(1-F(M)) \Big\}\,.$$
Further, by log-concavity property of $F$ and $1-F$, we have $\ell_M > 0$.

We also let $B = \max_{v} f(v)$ and $B' = \max_{v} f'(v)$, respectively denote the maximum value of the density function $f$ and the its derivative $f'$.

The following theorem bounds the regret of our PSGD policy.

\begin{theorem}\label{thm1}
Consider model~\eqref{eq:model} for the product market values and let Assumption~\ref{ass1} hold. Set $M = 2\l1u+\varphi^{-1}(0)$, with $\varphi$ being the virtual valuation function w.r.t distribution $F$. 
Then, the regret of PSGD pricing policy using a non-increasing sequence of step sizes $\{\eta_t\}_{t\ge1}$ is bounded as follows:
\begin{align}\label{thm:regret1}
\Reg(T) \le \frac{2(2B+MB')}{\ell_M} \max\bigg\{\frac{16}{\ell_M} \log T,
\,\frac{2\l1u^2}{\eta_{T+1}} + \frac{u_M^2}{2} \sum_{t=1}^T \eta_t + 2\l1u \sum_{t=1}^T  \frac{\delta_t}{\eta_t} \bigg\}+\frac{M}{T}\,,
\end{align}
where $\delta_t \equiv \|\th_{t+1}-\th_t\|$. 
\end{theorem}

In particular, if $\eta_t \propto \frac{1}{\sqrt{t}}$, then there exists a constant $C = C(B,M,W,\ell_M, u_M) >0$, independent of $T$, such that
\begin{align}
\Reg(T) \le C\Big(\sqrt{T} + \sum_{t=1}^T \sqrt{t} {\delta_t}\Big)\,.
\end{align}

At the core of our regret analysis (proof of Theorem~\ref{thm1}) is the following Lemma that provides a prediction error bound for the customer's valuations.
\begin{lemma}\label{lem:PE}
Consider model~\eqref{eq:model} for the product market values and let Assumption~\ref{ass1} hold.
Set $M = 2\l1u+\varphi^{-1}(0)$, with $\varphi$ being the virtual valuation function w.r.t distribution $F$.  
Let $\{\hth_t\}_{t\ge1}$ be generated by PSGC pricing policy, using a non-increasing positive series $\eta_{t+1}\le \eta_t$. Then, with probability at least $1-\frac{1}{T^2}$ the following holds true:
\begin{align}
\sum_{t=1}^T\<x_t,\th_t-\hth_t\>^2
&\le \frac{4}{\ell_M} \max\bigg\{\frac{16}{\ell_M} \log T, \nonumber\\
&\quad \quad \quad\quad\quad\;\;  \frac{2\l1u^2}{\eta_1} + \sum_{t=1}^T \Big(\frac{1}{2\eta_{t+1}} - \frac{1}{2\eta_t} \Big) \|\th_{t+1}-\hth_{t+1}\|^2  
+\frac{u_M^2}{2} \sum_{t=1}^T \eta_t + 2\l1u \sum_{t=1}^T  \frac{\delta_t}{\eta_t} \bigg\}\,,\label{eq:PE}
\end{align}
 where $u_M, \ell_M$  are given by Equations~\eqref{eq:uM}, \eqref{eq:lM}, respectively.
\end{lemma}
Lemma~\ref{lem:PE} is presented in a form that can also be used in proving our next results under the stochastic features model.
For proving Theorem~\ref{thm1}, we simplify bound~\eqref{eq:PE} as follows. Given that $\th_{t+1},\hth_{t+1}\in \Theta$, we have $\|\th_{t+1}-\hth_{t+1}\|\le 2\l1u$. Using the non-increasing property 
of sequence $\eta_t$, we write
\begin{align*}
\frac{2\l1u^2}{\eta_1} + \sum_{t=1}^T \left(\frac{1}{2\eta_{t+1}} - \frac{1}{2\eta_{t}}\right) \|\th_{t+1}-\hth_{t+1}\|^2 \le \frac{2\l1u^2}{\eta_1}  +  \sum_{t=1}^T \left(\frac{2\l1u^2}{\eta_{t+1}} - \frac{2\l1u^2}{\eta_{t}}\right) 
\le \frac{2\l1u^2}{\eta_{T+1}}\,.
\end{align*}
Therefore, bound~\eqref{eq:PE} simplifies to:
\begin{align}
\sum_{t=1}^T\<x_t,\th_t-\hth_t\>^2
\le \frac{4}{\ell_M} \max\bigg\{\frac{16}{\ell_M} \log T, 
\frac{2\l1u^2}{\eta_{T+1}}   
+\frac{u_M^2}{2} \sum_{t=1}^T \eta_t + 2\l1u \sum_{t=1}^T  \frac{\delta_t}{\eta_t} \bigg\}\,,\label{eq:PE1}
\end{align}
The regret bound~\eqref{thm:regret1} is derived by relating regret at each time period to the prediction error at that time. We refer to 
Section~\ref{sec:thm} for the proof of Theorem~\ref{thm1}.
 
\begin{remark}
The regret bound~\eqref{thm:regret1} does not depend on the dimension $d$, which makes our pricing policy desirable for high-dimensional applications.
Also, note that the temporal variation $\delta_t$ appears in our bound with coefficient $\sqrt{t}$. Therefore, variations at later times are more impactful on the regret of PSGD pricing policy.  
This is expected because at later times, the pricing policy is more relied on the accumulated information about the valuation model and an abrupt change in the model parameters can make
this information worthless. On the other side, temporal changes at the beginning steps are not that effective since the policy is still experimenting different prices to learn the customer's behavior.    
\end{remark}
\begin{remark}
While the regret bound is dimension-free, the computational complexity of PSGD pricing policy  scales with dimension $d$. Specifically, the complexity of each step is $O(d)$. To see this, we note that the gradient $\nabla \ell_t(\th)$
can be computed in $O(d)$ by Equations~\eqref{eq:nabla-nabla2} and~\eqref{eq:mu}. Projection onto set $\Theta$ ($\ell_2$ projection) is also $O(d)$.
\end{remark}
%==============================
\section{Stochastic features model}\label{stochastic}
In Theorem~\ref{thm1}, we showed that our PSGD pricing policy achieves regret of order $O(\sqrt{T}+\sum_{t=1}^T \sqrt{t} \delta_t)$. Let us point out that in Theorem~\ref{thm1} 
the arrivals (feature vectors $x_t$) are modeled as adversarial. In this section, we assume that features $x_t$ are independent and identically distributed according to 
a probability distribution on $\reals^d$. Under such stochastic model, we show that the regret of PSGD pricing scales at most of order $O(d^2\log T+\sum_{t=1}^T t\delta_t/d)$.

We proceed by formally defining the stochastic features model.
\begin{assumption}\label{SMF} (\emph{Stochastic features model}). Feature vectors $x_t$ are generated independently according to a probability distribution $\prob_{\bx}$, with a bounded support in $\reals^d$.
We denote by $\Sigma$ the covariance matrix of distribution $\prob_{\bx}$ and assume that $\Sigma$ has bounded eigenvalues. Specifically, there exist constants $C_{\min}$ and $C_{\max}$
such that for every eigenvalue $\sigma$ of $\Sigma$, we have $0<\frac{1}{d}C_{\min} \le \sigma < \frac{1}{d}C_{\max}$. 
\end{assumption}

 Without loss of generality and to simplify the presentation, we assume that $\prob_{\bx}$ is supported on the unit $\ell_2$ ball in $\reals^d$.
The rationale behind the above assumption on the scaling of eigenvalues is that $\Tr(\Sigma) = \E(\|x_t\|^2)\le 1$. Therefore, the assumption above on the eigenvalues of $\Sigma$  states that all the eigenvalues are of the same order.
 
%The above assumption holds for many common probability distributions, such as uniform, truncated normal, and in general truncated version of many more distributions. Generally, if $\pX$ is bounded below from zero on an open set around the origin, then it has a positive definite covariance matrix.

Under the stochastic features model, we define the notion of worst-case regret as follows.
For a policy $\pi$ be the seller's policy that sets price $p_t$ at period $t$, the $T$-period regret is defined as:
\begin{align}\label{regret-2}
\Reg^\pi(T) \equiv \sup\, \left\{\Delta^\pi_{\bth,\prob_{\bx}}:\, \th_t\in \Theta, \, \prob_{\bx} \in Q \right\}\,,
\end{align}
where  $Q$ denotes the set of probability distribution supported on $\ell_2$ unit ball satisfying Assumption~\ref{SMF} (bounded eigenvalues). Further, for $T\ge 1$, $\bth= (\th_1,\dotsc, \th_T)$ and probability measure $\prob_{\bx}$, we define
\begin{align}
 \Delta^\pi_{\bth,\prob_{\bx}}(T) = \E_{\bth,\prob_{\bx}} \left[\sum_{t=1}^T \bigg(p^*_t \ind(v_t \ge p^*_t) - p_t \ind(v_t \ge p_t) \bigg)\right]\,.
\end{align}
where the expectation is with respect to the distributions of idiosyncratic noise, $z_t$, and $\prob_{\bx}$, the distribution of feature vectors. 
Note the subtle difference with definition~\eqref{eq:Regret_def}, in that the worst case is computed over $Q$ rather than $\cX$. 

We propose a similar PSGD pricing policy for this setting, with a specific choice of the step sizes. Ideally, we want to set $\eta_t  = 6/(\ell_M C t)$, where $C$ is an arbitrary fixed constant such that $0<C<\sigma_{\min}$, with $\sigma_{\min}$ being the minimum eigenvalue of population covariance $\Sigma$. Of course, $\Sigma$ is unknown and therefore we  proceed as follows. We let $Q_t = (1/t)\sum_{\ell=1}^t x_\ell x_\ell^\sT$ be the empirical covariance based on the first $t$ features.  Denote by $\sigma_t$ the minimum eigenvalue of $Q_t$. We then use the sequence $\sigma_t$, and set the step size $\eta_t$ as 
\[\eta_t = \frac{1}{\lambda_t \cdot t}\,,\quad \quad \lambda_t = \frac{\ell_M}{6} \left\{\frac{1}{t}\Big(1+\sum_{\ell=1}^t \sigma_\ell\Big) \right\}\,.\]
Description of the PSGD pricing policy is given in Table above.

 \subsection{Logarithmic regret bound}
\begin{algorithm}[t]
\caption*{{\bf PSGD pricing policy for stochastic features model}}\label{alg-SFM}
\begin{algorithmic}[1]

\REQUIRE{\bf (at time $0$)} function $g$, set $\Theta$,\hspace{2cm}
\REQUIRE{\bf (arrives over time)} covariate vectors $\{x_t\}_{t\in \naturals}$ 

\ENSURE prices $\{p_t\}_{t\in \naturals}$ 

%\hnx{Is is possible to drop the line numbers?}
\STATE $p_1 \leftarrow 0$ and initialize $\hth_1 \in \Theta$
\STATE $Q_1 \leftarrow x_1x_1^\sT$

\FOR{$t = 1,2,3,\dots$}
\STATE Define $\sig_t$ as the minimum eigenvalue of $Q_t$.
\STATE Set 
\begin{equation}\label{lambdat}
\lambda_t=
%\begin{cases}
%1& \text{ if } \sig_t = 0\\
%\min(\frac{1}{2}\sig_t,1) & \text{ otherwise}
%\end{cases}
\frac{\ell_M}{6t}(1+\sum_{\ell=1}^t \sig_\ell)\,.
\end{equation}
\STATE Set 
\begin{align}\label{lambdat2}
\eta_t = \frac{1}{\lambda_t \cdot t}
\end{align}
\STATE Set $\hth_{t+1}$ according to the following rule:
\begin{eqnarray}\label{eq:ML}
\hth_{t+1} = \Pi_\Theta(\hth_t - \eta_t \nabla\ell_t(\hth_t)) 
\end{eqnarray}
with 
\begin{align} \label{eq:lt}
\ell_t(\th) = - \ind(y_t =1) \log (1-F(p_t - \<x_t,\th\> )) - \ind(y_t =-1) \log (F(p_t - \<x_t,\th\> )) 
\end{align}
%
%\begin{align} \label{eq:log_likelihood}
%\cL(\th) = - \frac{1}{\tau_{k-1}}\sum_{t\in \cT_{k-1}} \bigg\{\ind(y_t =1) \log (1-F(p_t - \th\cdot x_t )) + \ind(y_t =-1) \log (F(p_t - \th\cdot x_t )) \bigg\}\,.
%\end{align}
%
\STATE $Q_{t+1}\leftarrow (\frac{t}{t+1}) Q_t + (\frac{1}{t+1}) x_{t+1}x_{t+1}^\sT$
\STATE Set price $p_{t+1}$ as
\begin{align}\label{eq:price}
p_{t+1} \leftarrow g( \<x_{t+1},\hth_{t+1} \>) 
\end{align}

\ENDFOR
\end{algorithmic}
\end{algorithm}

The following theorem bounds the regret of our dynamics pricing policy.
\begin{theorem}\label{thm:log-regret}
Consider model~\eqref{eq:model} for the product market values and suppose Assumption~\ref{ass1} holds. Let $M = 2\l1u+\varphi^{-1}(0)$, with $\varphi$ being the virtual valuation function w.r.t distribution $F$. 
Under the stochastic features model (Assumption~\ref{SMF}), the regret of PSGD pricing policy is bounded as follows:
\begin{align}\label{eq:log-term}
\Reg(T) \le C_1 d^2 \log T + C_2\sum_{t=1}^T \frac{t}{d}\delta_t\,,
\end{align}
where $\delta_t \equiv \|\th_{t+1}-\th_t\|$ and $C_1, C_2$ are constants that depend on $C_{\max}, C_{\min}, u_M, \ell_M, M, B, \l1u$ but are independent of dimension $d$. 
\end{theorem}
 Proof of Theorem~\ref{thm:log-regret} relies on the following lemma that is analogous to Lemma~\ref{lem:PE} and establishes a prediction error bound for the customer's valuations.
\begin{lemma}\label{lem:PE-random}
Consider model~\eqref{eq:model} for the product market values and the stochastic features model (Assumption~\ref{SMF}).
Suppose that Assumption~\ref{ass1} holds and set $M = 2\l1u+\varphi^{-1}(0)$, with $\varphi$ being the virtual valuation function w.r.t distribution $F$. 
Let $\{\hth_t\}_{t\ge 1}$ be generated by PSGD pricing policy. Then,  
\begin{align*}
{C_{\min}}\sum_{t=1}^T  \E(\|\th_t-\hth_t\|^2)  \le&
\left[\frac{128}{\ell_M^2} + \frac{24u_M^2}{\ell_M^2} \left(\tilde{c}+ \frac{4}{C_{\min} d} \right) \right]\cdot d^3 \log T\\
& +8\l1u^2d \left(\frac{1}{T} +\frac{12}{\ell_M^2} + \frac{1}{c_2 d}\right) +{4\l1u}\sum_{t=1}^T t\delta_t\,.
\end{align*}
Here $\sigma_{\min}$ denotes the minimum eigenvalue of covariance $\Sigma$. (See Assumption~\ref{SMF}.)
\end{lemma}
%
%%%%%%%%%%%%%%%%%%%%%%%%%%%%%%%%%%%%%%%%%%%%%%%%%%%%%%%%%%%%%%
\subsection{A lower bound on regret}
In this section, we provide a theoretical lower bound on the minimum achievable regret of any pricing policy under the stochastic features model. Prior to that, we need to adopt a few notations. 

For a given time horizon $T$ and a sequence of valuations parameters $\bth = (\th_1, \dotsc, \th_T)$, let 
\begin{align}\label{V:T}
V_{\bth}(T) \equiv \sum_{t=1}^T t \|\th_{t+1}-\th_t\|\,. 
\end{align}
We also define, for $\nu\in[1/2,2]$,
\begin{align}
\cV(T,B,\nu) \equiv \{\bth:\, \th_t\in \Theta,\, V_{\bth}(T)\le B d  T^\nu \}\,.
\end{align}
By assuming $\bth\in \cV(T,B,\nu)$ for all $T$, we are assuming that nature has a finite temporal variation budget to use in changing the valuation parameters throughout the time horizon. Of course, different variation metrics can be considered such as total variation $\sum_{t=1}^T\delta_t$ or the maximum temporal variation $\sup_{1\le t \le T} \delta_t$ and the performance of a pricing policy can be studied under different variation budget constraints. The specific choice of~\eqref{V:T} is putting higher weights at later variations in the sequence $\bth$ and is reasonable for applications where one expects the buyer's preferences (valuation parameters) become stable over time. Note that designing favorable pricing policy for applications with gradual changes in buyer's preferences is more challenging than that for environments with bursty changes. This might look counterintuitive at first glance because at any time, the accumulated information about valuations can become useless by an abrupt change in the valuation model. However, as noticed and analyzed in~\cite{Keskin-TVC}, 
this is not that case because, intuitively, gradual changes can be undetectable and lead to significant revenue loss, while for bursty changes, the policy can be designed in a way to detect the changes and reset its estimate of the valuation model after each change to avoid large estimation error and revenue loss.
For a pricing policy $\pi$, consider the $T$-period regret, defined as
\begin{align}
\Reg^\pi(T,B,\nu) \equiv \max\Big\{\Delta^{\pi}_{\bth,\prob_{\bx}}(T): \, \bth\in \cV(T,B,\nu),\, \prob_{\bx} \in Q \Big\} 
\end{align}
where we recall that
\begin{align}\label{Rtth}
\Delta^{\pi}_{\bth,\prob_{\bx}}(T) \equiv \sum_{t=1}^T \E_{\bth,\prob_{\bx}}\bigg(p^*_t \ind(v_t \ge p^*_t) - p_t \ind(v_t \ge p_t) \bigg)\,.
\end{align}
Note that this is the same regret notion defined in~\eqref{regret-2}, where we just make the variation budget constraint explicit in the notation.

Rephrasing the statement of Theorem~\ref{thm:log-regret}, for PSGD pricing policy we have $\Reg^\pi(T,B,\nu) \le  C_1d^2 \log T + C_2 B T^\nu$. We next provide a lower bound on the regret of any pricing policy.
Indeed this lower bound applies to a powerful clairvoyant who fully observes the market values after the price is
either accepted or rejected.
\begin{theorem}\label{thm:LB-reg}
Consider linear model~\eqref{eq:model} where the market values $v_t(x_t)$, $1\le t \le T$, are fully
observed. We further assume that market value noises are generated as $z_t \sim \normal(0, \sigma^2)$.
There exists a constant $c$, depending on $\sigma$, $C_{\max}$, such that $\Reg^\pi(T,B,\nu) \ge c\min\Big(\left({B^2 dT^{2\nu-1}} \right)^{1/3}, T/d\Big)$, for any pricing policy $\pi$ and time horizon $T$.
\end{theorem}

The high-level intuition behind this result is that the nature can change the valuation parameters in a gradual manner such that the seller should pay a revenue loss in order to detect the changes and learn the new valuation parameter after a change. To be more specific, we divide the time horizon into cycles of length $N$ periods, where $N$ is of order $(T^{4-2\nu}/d)^{1/3}$ and consider a setting where the value of $\th_t$ can change to one of two options $\th^0$, $\th^1$, only in the first period of a cycle. We choose the parameter change $\delta = \|\th^1-\th^0\|$ of order $\sqrt{d/N}$ to ensure that $(i)$ no policy can identify the change without incurring a revenue loss of order $N\delta^2/d$ $(ii)$ The variation metric $V_{\bth}(T)$ remains below the allowable limit of $Bd T^\nu$. Therefore, the total regret over $T$ periods works out at $T\delta^2/d$. In particular, for proving point $(i)$ we quantify the likelihood of valuations under the probability measures corresponding to $\th^0$ and $\th^1$, using Kullback-Leibler divergence. We use Pinsker inequality form probability theory and hypothesis testing results from information theory to show that there is a significant probability of not detecting the (potential) change, which consequently yields a revenue loss of order $N\delta^2/d$, over each cycle. 

We refer to Section~\ref{proof:LB-reg} for the proof of Theorem~\ref{thm:LB-reg}. 
%%%%%%%%%%%%%%%%%%%%%%%%%%%%%%%%%%%%%%%%%%%%%%%%%%%%%%%%%%%%%%
\section{Numerical experiments}\label{sec:numerical}
We numerically study the performance of our PSGD pricing policy on synthetic data.
In our experiments, we set $\l1u = 5$ and set $\theta_1 = (\l1u/2) (Z/\|Z\|)$, with $Z\sim \normal(0,\id_d)$ a multivariate normal variable. We then generate a sequence of parameters $\theta_t$ as follows:
$$\theta_{t+1} = \th_t + r_t\,,$$
where $r_t = t^{-b} (\tZ/\|\tZ\|)$, with $\tZ\sim \normal(0,\id_d)$. Note that $\delta_t = \|\theta_{t+1}-\theta_t\| = \|r_t\| = t^{-b}$. 

Next, at each time $t$, product covariates $x_t$ are independently sampled from a Gaussian distribution $\normal(0,\id_d)$ and normalized so that $\|x_t\| = 1$. Further, the market shocks are generated as $z_t \sim \normal(0,\sigma^2)$, with $\sigma = 1$. We run the PSGD pricing policy for stochastic features model.
\bigskip

\begin{figure}[]
    \centering
        \includegraphics[width = 8cm]{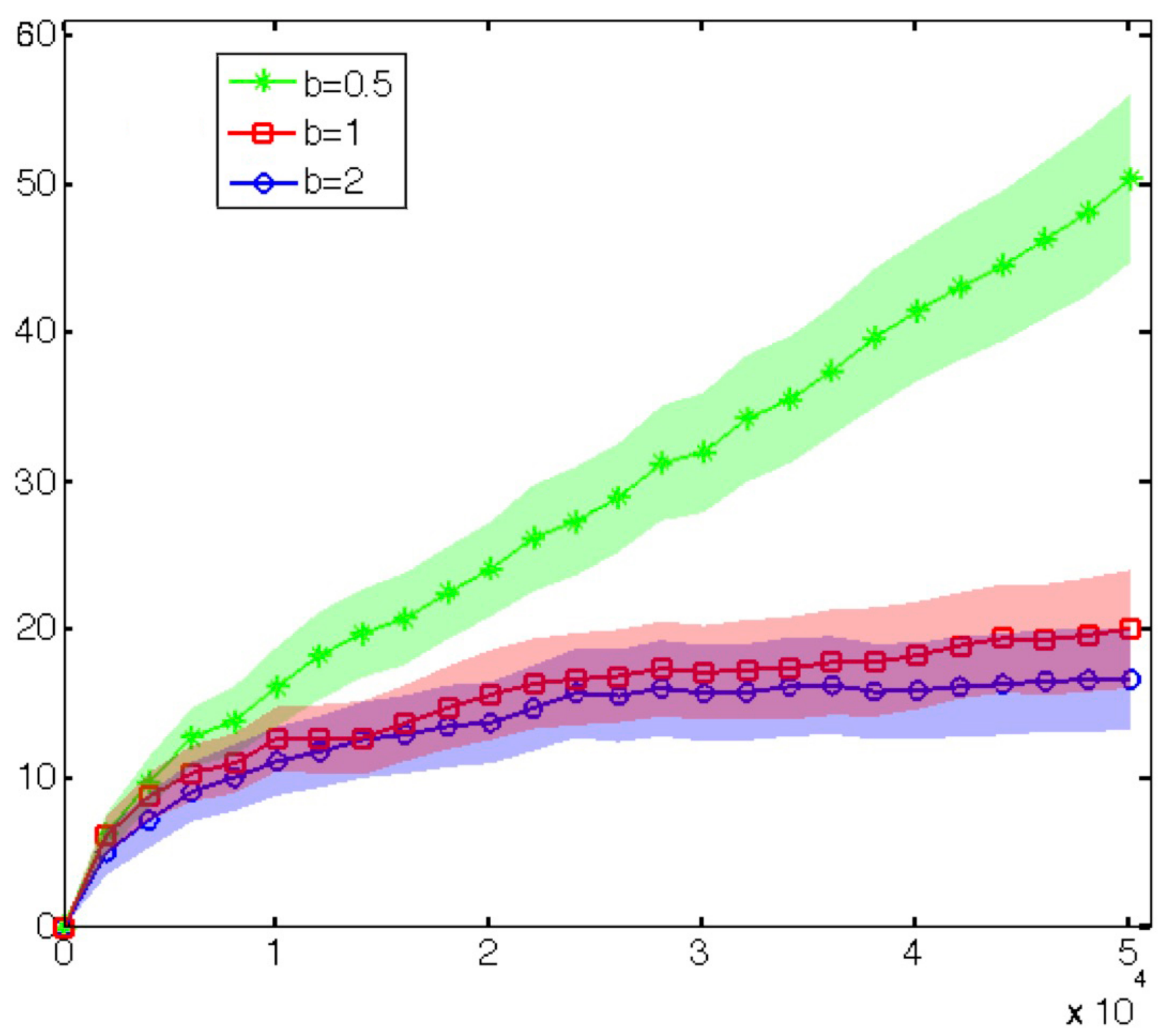}
        \put(-120,0){$T$}
        \put(-240,95){\rotatebox{90}{$\Reg$}}
    \caption{Cumulative regret of PSGD pricing policy for the synthetic data in Section~\ref{sec:numerical}. Temporal variations are $\delta_t  = t^{-b}$ and the curves are obtained by averaging across $80$ trials. Shaded region around each curve is the $95\%$ confidence interval.}
    \label{fig:regret_b}
\end{figure}

\noindent{\bf Results.} Figure~\ref{fig:regret_b} compares the cumulative regret (averaged over 80 trials) of the PSGD policy, for $b = 0.5, 1, 2$, on the aforementioned synthetic data for $T = 50,000$ steps. The shaded region around each curve correspond to the $95\%$ confidence interval across the $80$ trials. As expected, increase in $b$ results in larger temporal variations and larger regret.

\begin{figure}[]
    \centering
    \subfigure[$b = 0.5$]{
        \includegraphics[width = 5cm]{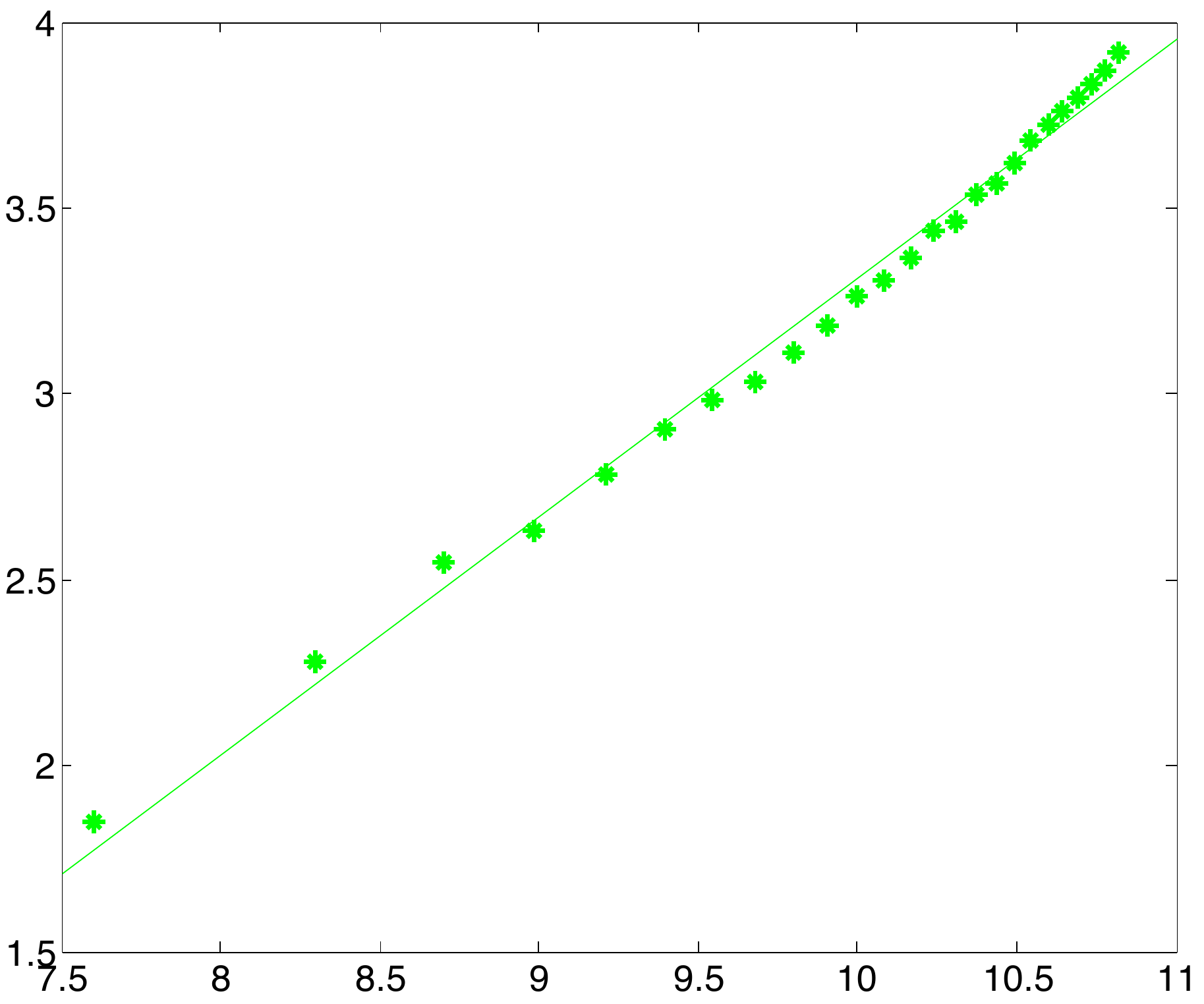}
        \put(-83,-10){{\small $\log (T)$}}
         \put(-83,-20){\phantom{AAA}}
           \put(-156,40){\rotatebox{90}{{\small $\log(\Reg)$}}}
        \label{fig:b05}
        }\hspace{1cm}
    \subfigure[$b = 1$]{
        \includegraphics[width=5cm]{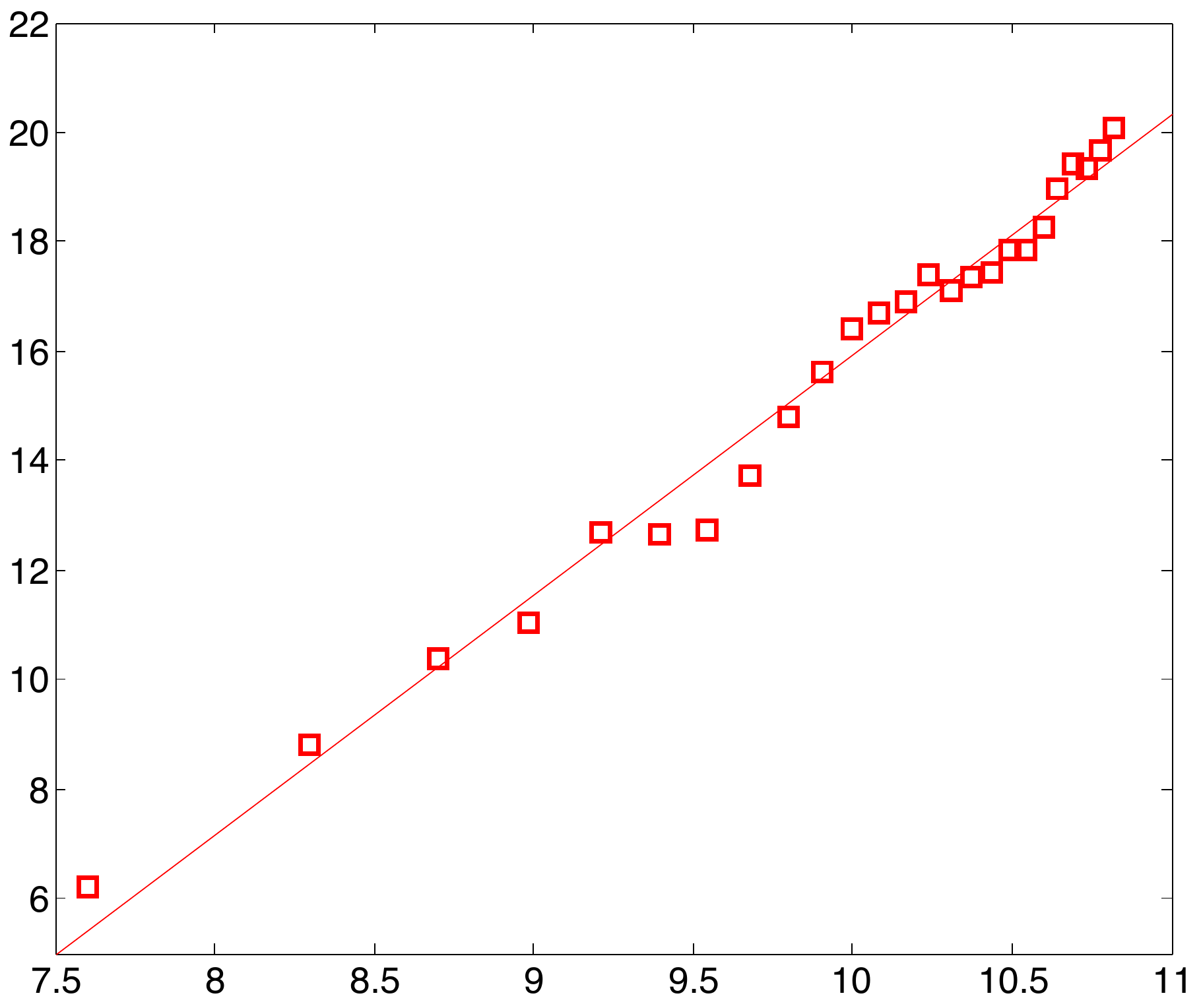}
         \put(-83,-10){{\small $\log (T)$}}
         \put(-83,-20){\phantom{AAA}}
           \put(-156,50){\rotatebox{90}{{\small $\Reg$}}}
        \label{fig:b1}
        }\hspace{0.3cm}
     \subfigure[$b = 2$]{
        \includegraphics[width=5cm]{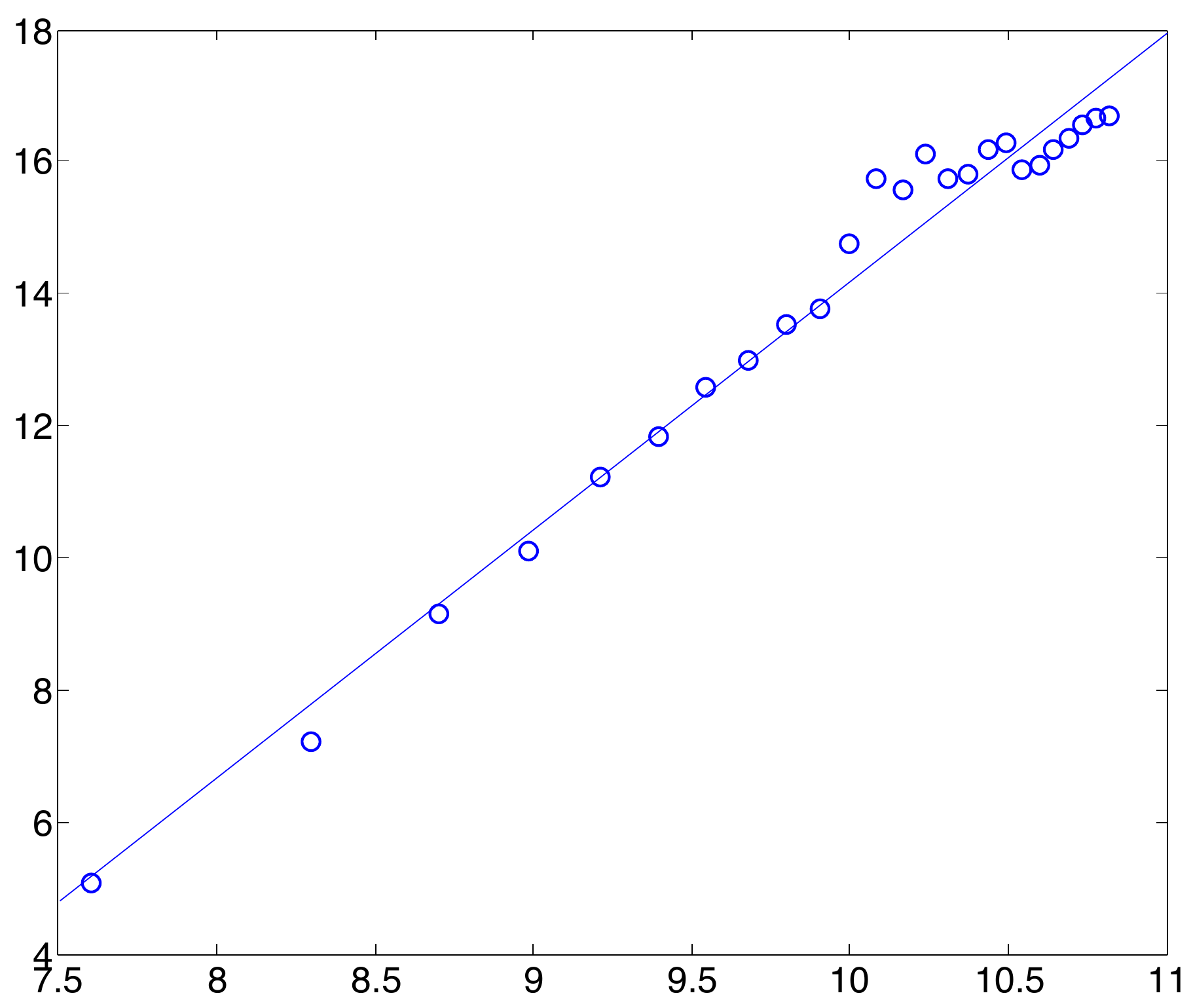}
        \put(-83,-10){{\small $\log (T)$}}
         \put(-83,-20){\phantom{AAA}}
           \put(-156,50){\rotatebox{90}{{\small $\Reg$}}}
        \label{fig:b2}
        }
    \caption{Cumulative regrets of PSGD for different  values of $b$. For $b=0.5$, $\Reg(T)\sim T^{2/3}$; for $b=1,2$, $\Reg(T)\sim \log(T)$. }\label{fig:b}

\end{figure}
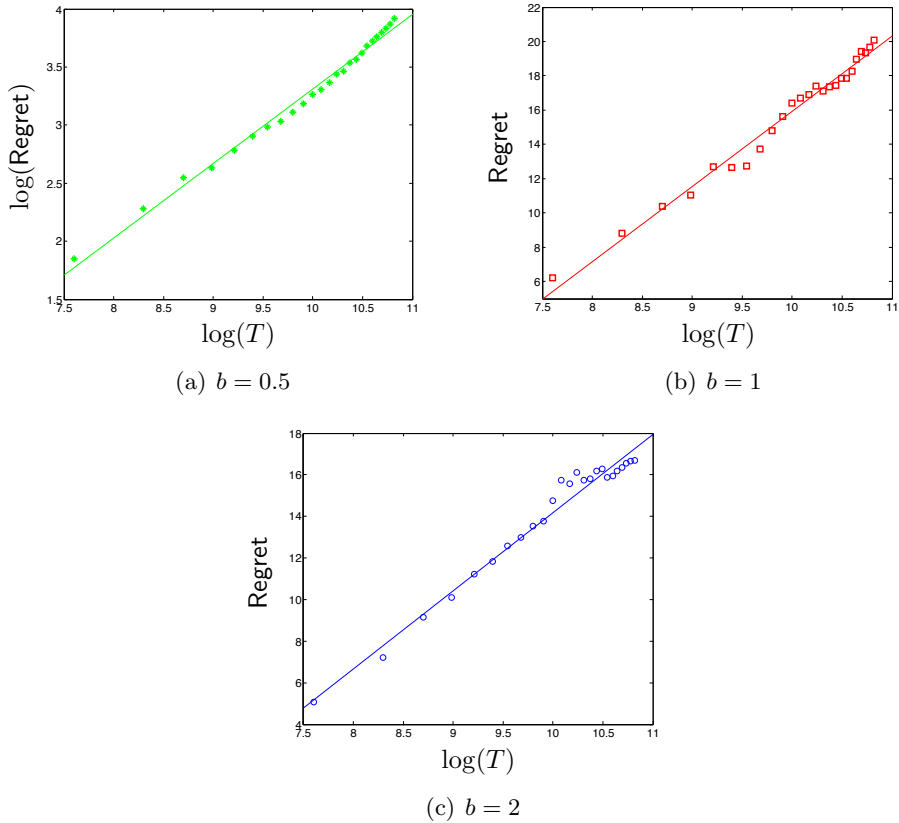

To better understand the behavior of regret for different values of $b$, we plotted the regret bounds in various scales in Figure~\ref{fig:b}. For $b= 0.5$, we have $\Reg(T)\sim T^{2/3}$, and for $b = 1, 2$, we have $\Reg~\sim \log(T)$. Comparing with Theorem~\ref{thm:log-regret}, we see that the empirical regret in case of $b=0.5$, $1$, is smaller than the upper bound given by Equation~\eqref{eq:log-term}, order-wise. However, it is worth noting that bound given in Theorem~\ref{thm:log-regret} applies to any adversarial choice of temporal variations $r_t$, while in our experiments we generated these terms independently at random. 

%
%%%%%%%%%%%%%%%%%%%%%%%%%%%%%%%%%%%%%%%%%%%%%%%%%%%%%%%%%%%%%%
\section{Extension to nonlinear model}\label{sec:extension}
Throughout the paper, we exclusively focused on linear models for buyer's valuation with varying coefficients. In order to generalize our results to nonlinear models, we consider a setting where the
market value of a product with feature vector $x_t$ is given by
\begin{align}
v_t(x_t) = \psi (\<x_t,\th_t\>+z_t)\,.
\end{align}
This model is often referred to as generalized linear model and captures nonlinear dependencies on features to some extent. 
We assume that the link function $\psi:\reals\mapsto \reals$ is a general log-concave function and is strictly increasing.

We next compute the pricing function. Since $\psi$ is strictly increasing, the expected revenue at a price $p$ amounts to $p\left(1-F\left(\psi^{-1}(p)-\<x_t,\th_t\>\right)\right)$. First order condition for the optimal price $p^*_t(x_t)$ reads as
\begin{align}\label{f6}
\psi'(\psi^{-1}(p_t^*)) = \frac{pf\left(\psi^{-1}(p_t^*)-\<x_t,\th_t\>\right)}{1-F\left(\psi^{-1}(p_t^*)-\<x_t,\th_t\>\right)}\,.
\end{align}
Define $\lambda(v) = f(v)/(1-F(v))$ the hazard rate function for distribution $F$, and let $\tp = \psi^{-1}(p)$. Writing~\eqref{f6} in terms of $\lambda$ function, we get
\begin{align}\label{f7}
\<x_t,\th_t\> = \tp^*_t - \lambda^{-1}\left(\frac{\psi'(\tp^*_t)}{\psi(\tp^*_t)}\right)\,.
\end{align}
For real-valued $v$, define 
\begin{align}\label{gpsi}
g_\psi^{-1}(v) \equiv v - \lambda^{-1} \left(\frac{\psi'(v)}{\psi(v)}\right)\,.
\end{align}  
Note that by log-concavity of $1-F$, the hazard function $\lambda$ is increasing. Also, by log-concavity of $\psi$, the term $\frac{\de}{\de v}\log\psi(v) = \psi'(v)/\psi(v)$ is decreasing.  Putting these together, we obtain that $ - \lambda^{-1} ({\psi'(v)}/{\psi(v)})$ is increasing. Therefore, the right-hand side of~\eqref{gpsi} is strictly increasing and the function $g_\psi$ is well-defined. Invoking Equation~\eqref{f7}, we derive the optimal price as
\begin{align}
p^*_t = \psi \left(g_\psi(\<x_t,\th_t\>)\right)\,.
\end{align}
As noted before, since $\psi$ is increasing, at each period $t$, a sale happens if $z_t\ge \psi^{-1}(p_t) - \<x_t,\th_t\>$. Hence, the log-likelihood function reads as
\begin{align}
\ell_t(\th) = - \ind(y_t =1) \log \left(1-F\left(\psi^{-1}(p_t) - \<x_t, \th\> \right)\right) - \ind(y_t =-1) \log \left(F\left(\psi^{-1}(p_t) - \<x_t, \th\> \right)\right)\,. 
\end{align}
In PSGD pricing policy, we run gradient step with this log-likelihood function and then set price $p_{t+1}$ at next step as $p_{t+1} = \psi \left(g_\psi(\<x_{t+1},\th_{t+1}\>)\right)$.

The results on the regret of PSGD pricing policy carries over to the generalized linear model as well. The analysis goes along the same lines and is omitted.
%%%%%%%%%%%%%%%%%%%%%%%%%%%%%%%%%%%%%%%%%%%%%%%%%%%%%%%%%%%%%%
\section{Proof of main theorems}\label{sec:thm}
\subsection{Proof of Theorem~\ref{thm1}}
\begin{lemma}\label{lem:M}
Set $M = 2\l1u+\varphi^{-1}(0)$, and for $\th\in \Theta$ define $u_t(\th) = p_t - \<x_t, \th\>$, where $p_t = g(\<x_t,\hth_t\>)$ is the posted price at time $t$. 
Then $|u_t(\th)|\le M$ for all $t\ge 1$.
\end{lemma}
Define function $h(;u)$ from $\reals_{\ge 0}$ to $\reals_{\ge 0}$ as 
\[h(p;u) = p(1-F(p-u))\]
This is the expected revenue at price $p$ when the noiseless valuation is $u$, i.e., $\<x_t,\th_t\> = u$. We let 
\begin{align}\label{Rt}
R_t \equiv p_t^* \ind(v_t\ge p_t^*) - p_t\ind(v_t\ge p_t)
\end{align}
be the regret incurred at time $t$, and define $\cF_t$ as the history up to time $t$ (Formally, $\cF_t$ is the $\sigma$-algebra generated by market noise $\{z_\ell\}_{\ell=1}^{t}$.)
Then,
\begin{align}
\E(R_t|\cF_{t-1}) = p_t^* \prob(v_t\ge p_t^*) - p_t \prob(v_t\ge p_t) = h(p^*_t;\<x_t,\th_t\>) - h(p_t;\<x_t,\hth_t\>)\,.
\end{align}
The optimal price $p_t^*$ is the maximizer of $h(p;\<x_t,\th_t\>)$ and thus $h'(p_t^*;\<x_t,\th_t\>) = 0$. By Taylor expansion of function $h$, there exists a value $p$ between $p_t$ and $p^*_t$, such that, 
\begin{align}\label{h-taylor}
h(p_t;\<x_t,\th_t\>) -  h(p^*_t;\<x_t,\th_t\>)= \frac{1}{2} h''(p;\<x_t,\th_t\>) (p_t-p^*_t)^2\,.
\end{align}

We next show that $|h''(p;\<x_t,\th_t\>)|\le C$ with $C = 2B+MB'$. Recall that $B = \max_v f(v)$ and $B'= \max_v f'(v)$. To see this, we write
\begin{align}\label{h"}
|h''(p;\<x_t,\th_t\>)| = \Big|2f(p-\<x_t,\th_t\>)+pf'(p-\<x_t,\th_t\>)\Big| \le 2B+MB'\,.
\end{align}
Putting Equations~\eqref{Rt}, \eqref{h-taylor}, \eqref{h"} and using the $1$-Lipschitz property of price function $g$, we conclude:
\begin{align}\label{eq:chain}
\E[R_t|\cF_{t-1}] &= h(p^*_t;\<x_t,\th_t\>) - h(p_t;\<x_t,\hth_t\>)\le \frac{2B+MB'}{2}(p_t-p^*_t)^2\nonumber\\
&= \frac{2B+MB'}{2}\Big(g(\<x_t,\hth_t\>)-g(\<x_t,\th_t\>)\Big)^2 \le \frac{2B+MB'}{2} \<x_t,\th_t-\hth_t\>^2
\end{align}
To ease the presentation, define the shorthand
\[
A(T) \equiv \frac{4}{\ell_M} \max\bigg\{\frac{16}{\ell_M} \log T,
\,\frac{2\l1u^2}{\eta_{T+1}} + \frac{u_M^2}{2} \sum_{t=1}^T \eta_t + 2\l1u \sum_{t=1}^T  \frac{\delta_t}{\eta_t} \bigg\}\,.
\]
We further let $\cG$ be the probabilistic event that $\sum_{t=1}^T \<x_t,\th_t-\hth_t\>^2\le A(T)$.  Employing Lemma~\ref{lem:PE} and using the fact that $\|\th_{t+1}-\hth_{t+1}\|^2\le 4\l1u^2$, we obtain that
$\prob(\cG) \ge 1-\frac{1}{T^2}$.

We continue by bounding $E(R_t)$ as follows:
\begin{align*}
\E[R_t] &= \E[\E[R_t|\cF_{t-1}]] = \E\Big[\E[R_t|\cF_{t-1}]\cdot \Big(\ind(\cG)+\ind(\cG^c)\Big)\Big]  \\
&= \frac{2B+MB'}{2} \E\Big[\<x_t,\th_t-\hth_t\>^2 \cdot \ind(\cG)\Big] + M\prob(\cG^c)\,.
\end{align*}
Consequently,
\begin{align*}
\Reg(T) \le \sum_{t=1}^T \E[R_t] &\le \frac{2B+MB'}{2}\E\Big[\sum_{t=1}^T \<x_t,\th_t-\hth_t\>^2 \cdot \ind(\cG)\Big] + M T\, \prob(\cG^c)
\le \frac{2B+MB'}{2} A(T) + \frac{M}{T}\,.
\end{align*}
 The proof is complete.
 %============
 \subsection{Proof of Theorem~\ref{thm:log-regret}}
 Proof of Theorem~\ref{thm:log-regret} follows along the same lines as proof of Theorem~\ref{thm1}. Let $\ctF_t$ be the $\sigma$-algebra generated by market noises $\{z_\ell\}_{\ell=1}^t$ and feature vectors $\{x_{\ell}\}_{\ell=1}^{t}$. Further, let $\cF_t$ be the $\sigma$-algebra generated by $\ctF_t \cup\{x_{t+1}\}$.
 For term $R_t$ defined by~\eqref{Rt} and following the chain of inequalities as in~\eqref{eq:chain},
 \begin{align}
 \E[R_t|\cF_{t-1}] \le \frac{2B+MB'}{2}\, \<x_t,\th_t-\hth_t\>^2 \,.
 \end{align}
 For brevity in notation, let $\bar{B} = (2B+MB')/2$. 
 Since, $\cF_{t}\supseteq \ctF_t$, by iterated law of iteration,
 \begin{align}
\E(R_t|\ctF_{t-1}) = \E(\E(R_t|\cF_{t-1})|\ctF_{t-1}) \le \bar{B} \<\th_t-\hth_t, \Sigma (\th_t-\hth_t)\> \le \frac{1}{d} \bar{B}C_{\max} \|\th_t-\hth_t\|^2
 \end{align}
 Applying Lemma~\ref{lem:PE-random}, we get
 \begin{align*}
 \Reg(T) &\le \sum_{t=1}^T \E[R_t] \le \frac{1}{d}  \bar{B}C_{\max} \sum_{t=1}^T \E(\|\th_t-\hth_t\|^2)\\
 &\le \bar{B} \frac{C_{\max}}{C_{\min}}\left[\frac{128}{\ell_M^2} + \frac{24u_M^2}{\ell_M^2} \left(\tilde{c}+ \frac{4}{C_{\min} d} \right) \right]\cdot d^2 \log T\\
& +8\l1u^2\bar{B} \frac{C_{\max}}{C_{\min}}  \left(\frac{1}{T} +\frac{12}{\ell_M^2} + \frac{1}{c_2 d}\right) + \bar{B} \frac{C_{\max}}{C_{\min}}\Big(\frac{4\l1u}{d}\Big) \sum_{t=1}^T t\delta_t\,.
 \end{align*}
 The result follows by taking
 \begin{align*}
 C_1 &= \bar{B} \frac{C_{\max}}{C_{\min}} \left[{8\l1u^2}\left(\frac{1}{T} +\frac{12}{\ell_M^2} + \frac{1}{c_2 d}\right)
 + \frac{128}{\ell_M^2} + \frac{24u_M^2}{\ell_M^2} \left(\tilde{c}+ \frac{4}{C_{\min} d} \right)  \right]\,,\\
 C_2&= 4\l1u \bar{B} \frac{C_{\max}}{C_{\min}} \,.
 \end{align*}
 %
 
%=====================================
\subsection{Proof of Theorem~\ref{thm:LB-reg}}\label{proof:LB-reg}
The proof methodology is similar to the proof of~\cite[Theorem 1]{Keskin-TVC}.
 
We first propose a setting for constructing the sequence of valuation parameters $\bth= (\th_1,\dotsc,\th_T)$. Divide the time horizon into cycles of length $N = \lceil m_0 T^{(4-2\nu)/3}\rceil$, where $m_0 = (\frac{\sigma^2}{C_{\max}B^2d})^{1/3}$. Consider a setting wherein the noise markets are generated as $z_t\sim\normal(0,\sigma^2)$ and the value of $\th_t$ can change only in the first period of a cycle, taking one of the two values $\{\th^0,\th^1\}$.  Here, $\th^0, \th^1\in \reals^d$ are two arbitrary vectors such that $\|\th^0-\th^1\| =\delta$, with $\delta = \min(\sigma\sqrt{d/(C_{\max}N)},\sqrt{c_2})$. Note that for this sequence of $\bth$, we have
\begin{align}
V_{\bth}(T) \le \sum_{k=1}^{\lceil T/N\rceil} (kN) \delta \le\frac{T^2}{N} \delta \le Bd T^\nu
\end{align}
We consider a clairvoyant who fully observes the market values $v_t(x_t)$. Focus on a single cycle and let $\prob_0^\pi$ (resp. $\prob_1^\pi$) denote the probability distribution of valuations $(v_1, v_2, \dotsc, v_N)$ when all the parameters $\th_t $ are equal to $\th^0$ (resp. $\th^1$), for $1\le t \le N$. The KL divergence between $\prob_0^\pi$ and $\prob_1^\pi$ amounts to 
\begin{align}
\KL(\prob_0^\pi,\prob_1^\pi) \equiv \E_0^\pi \log \left(\frac{\prod_{t=1}^N \phi\left(\dfrac{v_t-\<x_t,\th_0\>}{\sigma}\right)}{\prod_{t=1}^N \phi\left(\dfrac{v_t-\<x_t,\th_1\>}{\sigma}\right)} \right)
\end{align}
where $\E_0^\pi$ denotes expectation w.r.t $\prob_0^\pi$ and $\phi(s) = 1/(\sqrt{2\pi})e^{-s^2/2}$ is the standard Gaussian density. After simple algebraic manipulation, we obtain
\begin{align*}
\KL(\prob_0^\pi,\prob_1^\pi) &= -\frac{1}{2\sigma^2} \E^\pi_0 \bigg\{\sum_{t=1}^N (2z_t - \<x_t,\th^1-\th^0\>)\<x_t,\th^1-\th^0\> \bigg\}\\
&= \frac{1}{2\sigma^2} \sum_{t=1}^N \E^\pi_0 (\<x_t,\th^1-\th^0\>^2) \le \frac{1}{2\sigma^2 d} \sum_{t=1}^N C_{\max} \|\th^1-\th^0\|^2\\
&= \frac{1}{2\sigma^2}C_{\max} \frac{\delta^2 N}{d}\,.
\end{align*}
We next relate the expected regret to the KL divergence between $\prob_0^\pi$ and $\prob_1^\pi$.

\begin{lemma}\label{lem:Rt-LB}
Let $R_t$ be the regret incurred at time $t$, defined as $R_t \equiv p_t^* \ind(v_t\ge p_t^*) - p_t\ind(v_t\ge p_t)$. Then, there exist constants $c_1, c_2$ depending on $\sigma$, $\l1u$, and $C_{\min}$, such that
\begin{align}
\E(R_t) \ge \frac{c_1}{d} \E\Big\{ \min\left(\|\hth_t-\th_t\|_2^2,c_2\right)\Big\}\,.
\end{align} 
\end{lemma}
Proof of Lemma~\ref{lem:Rt-LB} goes along the proof of~\cite[Equation (55)]{JavanmardNazerzadeh} and is omitted. 

By applying Lemma~\ref{lem:Rt-LB}, we have
 
\begin{align}
\Delta^\pi_{\bth,\prob_{\bx}}(N) = \sum_{t=1}^N \E_{\bth}(R_t) \ge \frac{c_1}{d} \sum_{t=1}^N \E\Big\{ \min\left(\|\hth_t-\th_t\|_2^2,c_2\right)\Big\}\,.
\end{align}

For brevity in notations, for the sequence $\bth = (\th_1,\dotsc, \th_N)$, we define $\dis_a(\bth) = c_1\sum_{t=1}^N \min(\|\th_t-\th^a\|_2^2,c_2)$, for $a=1,2$. 
Define two sets $J_a$, for $a=1,2$ as follows:
\begin{align}
J_a = \left\{\bth =(\th_1,\dotsc, \th_N):\, \th_i\in\reals^d\,,\, \dis_a(\bth) < \frac{1}{4} N \delta^2\right\}\,.
\end{align}
Then,

\begin{align*}
\max\left(\Delta^\pi_{0,\prob_{\bx}}(N),\Delta^\pi_{1,\prob_{\bx}}(N)\right) &\ge \frac{1}{d} \max\Big(\E_0^\pi (\dis_0(\bth)), \E^\pi_1(\dis_1(\bth)) \Big) \\
&\ge \frac{N}{4d} \delta^2 \max\Big(\prob^\pi_0(\bth\notin J_0), \prob^\pi_1(\bth\notin J_1) \Big)\\
&\stackrel{(a)}{\ge} \frac{N}{4d} \delta^2 \max\Big(\prob^\pi_0(\bth\notin J_0), \prob^\pi_1(\bth\in J_0) \Big)\\
&\ge \frac{N}{8d} \delta^2 \Big(\prob^\pi_0(\bth\notin J_0) + \prob^\pi_1(\bth\in J_0) \Big)\\
&\ge \frac{N}{8d} \delta^2 \Big(1- \prob^\pi_0(\bth\in J_0) + \prob^\pi_1(\bth\in J_0) \Big)\\
&\ge \frac{N}{8d} \delta^2 \Big(1- \sqrt{\frac{1}{2} \KL(\prob^\pi_0,\prob^\pi_1)} \Big) \quad \text{(By Pinsker inequality)}\\
&\ge  \frac{N}{8d} \delta^2 \left(1- \frac{1}{2\sigma} \delta \sqrt{C_{\max} \frac{N}{d}} \right) \ge \frac{N\delta^2}{16d} \,.
\end{align*}
Here $(a)$ holds because $\bth\in J_0$ implies $\bth\notin J_1$. Otherwise, $\dis_0(\bth)< N\delta^2/4$ and $\dis_1(\bth)< N\delta^2/4$. Using the inequality $\min(a + b, c) \le \min(a, c) + \min(b, c)$ for $a, b, c \ge 0$, and applying triangle inequality, we get
\begin{align}
N \min(\|\th^0-\th^1\|^2,c_2) \le 2\dis_0(\bth) + 2\dis_1(\bth) < N\delta^2\,,
\end{align}
which is a contradiction because $\delta^2 = \|\th^0-\th^1\|^2 \le c_2$.
Therefore, we conclude that 
\begin{align}
\Reg^\pi(T,B,\nu) &\ge \Big\lfloor \frac{T}{N} \Big\rfloor \max\left(\Delta^\pi_{0,\prob_{\bx}}(N),\Delta^\pi_{1,\prob_{\bx}}(N)\right) \nonumber\\
&\ge \frac{T\delta^2}{16d} = \frac{T}{16} \min\Big(\frac{\sigma^2}{C_{\max}N}, \frac{c_2}{d}\Big)\nonumber\\
&= \frac{1}{16} \min\left\{\Big(\frac{\sigma^2}{C_{\max}} \Big)^{2/3} ({B^2 dT^{2\nu-1}} )^{1/3}, \frac{c_2 T}{d}\right\}\,.
\end{align}
The result follows.
%======================================
\section{Proof of main lemmas}\label{sec:lem}
\subsection{Proof of Lemma~\ref{lem:PE}}
We prove Lemma~\ref{lem:PE} by developing an upper bound and a lower bound for the quantity $\sum_{t=1}^T \ell_t(\hth_t) - \sum_{t=1}^T \ell_t(\th_t)$.
The result follows by combining these two bounds.

\begin{lemma}[\emph{Upper bound}]\label{lem:DUB}
Suppose $\{\th_t\}_{t\ge 1}$ is an arbitrary sequence in $\Theta$, and $\|\th\|\le \l1u$ for all $\th\in \Theta$. 
Set $M = 2\l1u+\varphi^{-1}(0)$, with $\varphi$ being the virtual valuation function w.r.t distribution $F$.  
Further, let $\{\hth_t\}_{t\ge1}$ be generated by PSGD policy using a non-increasing positive series $\eta_{t+1}\le \eta_t$. Then
\begin{align}
\sum_{t=1}^T \ell_t(\hth_t) -\sum_{t=1}^T \ell_t(\th_t) \le &\frac{2\l1u^2}{\eta_{1}} + \sum_{t=1}^T \Big(\frac{1}{2\eta_{t+1}}-\frac{1}{2\eta_t}\Big)\|\th_{t+1}-\hth_{t+1}\|^2 \nonumber\\ 
&+\frac{u_M^2}{2} \sum_{t=1}^T \eta_t + 2\l1u \sum_{t=1}^T \frac{\delta_t }{\eta_t} -\frac{\ell_M}{2}\sum_{t=1}^T\<x_t,\th_t-\hth_t\>^2\,,\label{DUB}
\end{align}
where $\delta_t \equiv \|\th_{t+1}-\th_t\|$ and we recall $u_M$ from Equation~\eqref{eq:uM}.
\end{lemma}
The proof of Lemma~\ref{lem:DUB} uses similar ideas to the regret bounds established in~\cite{Hall-DGD}, but uses the log-concavity of $F$ and $1-F$ and also definition 
of $u_M$ and $\ell_M$ as per Equations~\eqref{eq:uM} and \eqref{eq:lM} to get a more refined bound including quadratic terms $\<x_t,\hth_t-\th_t\>^2$. We refer to Appendix~\ref{app:DUB}
for the proof of Lemma~\ref{lem:DUB}.

Our next Lemma provides a probabilistic lower bound on $\sum_{t=1}^T \ell_t(\hth_t) - \sum_{t=1}^T \ell_t(\th_t)$.
\begin{lemma}[\emph{Lower bound}]\label{lem:DLB}
Consider model~\eqref{eq:model} for the product market values and suppose Assumption~\ref{ass1} holds. 
Let $\{\hth_t\}_{t\ge 1}$ be an arbitrary sequence in $\Theta$. Then with probability at least $1-\frac{1}{T^2}$ the following holds true
\begin{align}\label{DLB}
\sum_{t=1}^T \ell_t(\hth_t) - \sum_{t=1}^T\ell_t(\th_t)  \ge -2\sqrt{\log T} \Big\{\sum_{t=1}^T \<x_t,\th_t-\hth_t\>^2\Big\}^{1/2} \,.
\end{align}
\end{lemma}
Proof of Lemma~\ref{lem:DLB} is given in Appendix~\ref{app:DLB}. It uses convexity of $\ell_t(\hth)$ and an application of a concentration bound on martingale difference sequences.

Combining Equations~\eqref{DUB} and \eqref{DLB} we obtain that with probability at least $1-\frac{1}{T^2}$ the following holds true
\begin{align}
-2\sqrt{\log T} \Big\{\sum_{t=1}^T \<x_t,\th_t-\hth_t\>^2\Big\}^{1/2}  \le& \frac{2\l1u^2}{\eta_1} + \sum_{t=1}^T \Big(\frac{1}{2\eta_{t+1}} - \frac{1}{2\eta_t} \Big) \|\th_{t+1}-\hth_{t+1}\|^2 \nonumber\\ 
&+\frac{u_M^2}{2} \sum_{t=1}^T \eta_t + 2\l1u \sum_{t=1}^T \frac{\delta_t }{\eta_t} -\frac{\ell_M}{2}\sum_{t=1}^T\<x_t,\th_t-\hth_t\>^2
\end{align}

Rearranging the terms, we get
\begin{align}
\frac{\ell_M}{2}&\sum_{t=1}^T\<x_t,\th_t-\hth_t\>^2 -2\sqrt{\log T} \Big\{\sum_{t=1}^T \<x_t,\th_t-\hth_t\>^2\Big\}^{1/2} \nonumber\\
&\le \frac{2\l1u^2}{\eta_1} + \sum_{t=1}^T \Big(\frac{1}{2\eta_{t+1}} - \frac{1}{2\eta_t} \Big) \|\th_{t+1}-\hth_{t+1}\|^2 
+\frac{u_M^2}{2} \sum_{t=1}^T \eta_t + 2\l1u \sum_{t=1}^T \frac{\delta_t }{\eta_t}\label{f4}
\end{align}
Define $A\equiv \sum_{t=1}^T\<x_t,\th_t-\hth_t\>^2 $ and denote by $B$ the right-hand side of Equation~\eqref{f4}. 

Writing in terms of $A$ and $B$, we have 
\begin{align}\label{f5}
 A - \frac{4}{\ell_M} \sqrt{A \log T} \le \frac{2B}{\ell_M}\,.
 \end{align}
We next upper bound $A$ as follows. Consider two cases:
\bigskip

{\bf Case 1:} Assume that 
\[
\sqrt{A\log T} \le \frac{\ell_M}{8}A\,.
\]
Using this in Equation~\eqref{f5}, we get $A\le 4B/\ell_M$.
\bigskip

{\bf Case 2:} Assume that 
\[
\sqrt{A\log T} > \frac{\ell_M}{8}A\,.
\]
Then, $A< (64/\ell_M^2)\log T$.

Combining the above two cases, we obtain 
$$A\le \frac{4}{\ell_M}\max\Big(\frac{16}{\ell_M}\log T, {B}\Big)\,.$$

Substituting for $A$ and $B$, we have
\begin{align*}
\sum_{t=1}^T\<x_t,\th_t-\hth_t\>^2 &\le \frac{4}{\ell_M}\max\bigg\{\frac{16}{\ell_M} \log T, \\
&\quad \quad \quad \quad\quad\; \;\frac{2\l1u^2}{\eta_1} + \sum_{t=1}^T \Big(\frac{1}{2\eta_{t+1}} - \frac{1}{2\eta_t} \Big) \|\th_{t+1}-\hth_{t+1}\|^2  
+\frac{u_M^2}{2} \sum_{t=1}^T \eta_t + 2\l1u \sum_{t=1}^T \frac{\delta_t }{\eta_t} \bigg\}
\end{align*}
The proof is complete.
   
%Below we provide an outline for the proof of Theorem~\ref{thm:regret} and defer its complete proof to Appendix~\ref{proof:thm2}.
%==================================================================
\subsection{Proof of Lemma~\ref{lem:PE-random}}\label{proof:PE-random}
\begin{propo}\label{pro:eigmin}
Let $\sig_t$ denote the minimum eigenvalue of $Q_t \equiv (1/t)\sum_{\ell=1}^t x_\ell x_\ell^\sT$. Further, let $\sigma_{\min}$ be the minimum eigenvalue of $\Sigma$, where $\Sigma$ is the population covariance of  feature vectors as in Assumption~\ref{SMF}. Then, there exist constants $c_1,c_2>0$, such that
\begin{align}
\forall t\ge c_1d:\,\, \prob\Big(\frac{1}{2} \sigma_{\min} \le \sig_t\le \frac{3}{2}\sigma_{\min}\Big) \ge 1- 2e^{-c_2 t/d}\,. \label{eq:prob}
\end{align}
Further, $\sig_t\le 1$, for all $t\ge 1$.
\end{propo}

Let $\cF_t$ be the $\sigma$ algebra generated by market shocks $\{z_\ell\}_{\ell=1}^t$ and features $\{x_\ell\}_{\ell=1}^t$. We further define $D_t = \<x_t,\hth_t-\th_t\>^2 - \|\Sigma^{1/2}(\hth_t-\th_t)\|^2$. Note that $\hth_t$ is $\cF_{t-1}$ measurable and $x_t$ is independent of $\cF_{t-1}$, which implies $\E(D_t|\cF_{t-1}) = 0$. Hence, $\E(D_t) = 0$ by iterated law of expectation and therefore $\sum_{t=1}^T \E(D_t) = 0$. Equivalently, 
\begin{align}\label{eq:PredictionB1}
\E\left[\sum_{t=1}^T \<x_t,\hth_t-\th_t\>^2 \right] = \sum_{t=1}^T \E\left[\|\Sigma^{1/2}(\hth_t-\th_t)\|^2 \right] \ge \sigma_{\min}\, \E\left[\sum_{t=1}^T \|\hth_t-\th_t\|^2\right]
\end{align}
Define $\cG_T$ the event that bound~\eqref{eq:PE} holds true.
Then,
\begin{align}
\E\left[\sum_{t=1}^T \<x_t,\hth_t-\th_t\>^2 \right] &= \E\left[\sum_{t=1}^T \<x_t,\hth_t-\th_t\>^2 \cdot(\ind_\cG+\ind_{\cG^c}) \right]\nonumber\\
&\le \E\left[\sum_{t=1}^T \<x_t,\hth_t-\th_t\>^2 \cdot\ind_\cG \right] + 4\l1u^2 T\, \prob(\cG^c)\nonumber\\
&\le \E\left[\sum_{t=1}^T \<x_t,\hth_t-\th_t\>^2 \cdot\ind_\cG \right] + \frac{4\l1u^2}{T}\,. 
 \label{eq:term0}
%&\le  {2\lambda\l1u^2}
%+ \E\left[\sum_{t=1}^T \lambda \|\th_t-\hth_t\|^2\right]+ \frac{u_M^2}{2\lambda}\log T +2\lambda \l1u \sum_{t=1}^T t\delta_t +4\l1u^2 \prob(G^c)\nonumber\\
% &\le  {C_{\min}\l1u^2}
% + \frac{C_{\min}}{2}\E\left[\sum_{t=1}^T \|\th_t-\hth_t\|^2\right]+ \frac{u_M^2}{2\lambda}\log T + C_{\min} \l1u \sum_{t=1}^T t\delta_t +\frac{4\l1u^2}{T^2} \label{eq:PredictionB2}
 \end{align}
Further, using inequality $\max(a,b) \le |a|+|b|$, we get
\begin{align}
\E\left[\sum_{t=1}^T \<x_t,\hth_t-\th_t\>^2 \cdot\ind_\cG \right] \le \frac{4}{\ell_M}\bigg\{&
\frac{16}{\ell_M}\log T + \frac{12\l1u^2}{\ell_M} + \frac{1}{2} \sum_{t=1}^T \E\bigg[\Big((t+1) \lambda_{t+1}-t \lambda_t \Big) \cdot \|\th_{t+1}-\hth_{t+1}\|^2\bigg]\nonumber\\
& + \frac{u_M^2}{2}\sum_{t=1}^T\E\left[\frac{1}{t\lambda_t}\right]+2\l1u \sum_{t=1}^T \E[t\lambda_t] \delta_t\bigg\}\,.\label{eq:PredictionB2}
\end{align}
We next bound the terms on the right-hand side individually.
\begin{align}
\sum_{t=1}^T &\E\bigg[\Big((t+1) \lambda_{t+1}-t \lambda_t \Big) \cdot \|\th_{t+1}-\hth_{t+1}\|^2\bigg] 
\le \frac{\ell_M}{6}\sum_{t=1}^T \E\bigg[\sig_{t+1} \cdot \|\th_{t+1}-\hth_{t+1}\|^2\bigg]\nonumber\\
&\le\frac{\ell_M}{6} \sum_{t=1}^T  \E\bigg[\sig_{t+1} \|\th_{t+1}-\hth_{t+1}\|^2\, \ind(\sig_{t+1}<3\sigma_{\min}/2)\bigg]+ \frac{\ell_M}{6} \sum_{t=1}^T  \E\bigg[\sig_{t+1} \, \|\th_{t+1}-\hth_{t+1}\|^2\, \ind(\sig_{t+1} > 3\sigma_{\min}/2)\bigg] \nonumber\\
&\le \frac{\ell_M}{4}\sigma_{\min}\sum_{t=1}^T  \E\Big(\|\th_{t+1}-\hth_{t+1}\|^2\Big) + \sum_{t=1}^T  2\ell_M\l1u^2 e^{-c_2 t/d}\nonumber\\
&\le  \frac{\ell_M}{4} \sigma_{\min}\sum_{t=1}^T  \E\Big(\|\th_{t+1}-\hth_{t+1}\|^2\Big) +  \frac{2\ell_M}{c_2d} \l1u^2 \,,\label{term1}
\end{align}
where in the last inequality, we used $\prob(\sig_{t+1} > 3\sigma_{\min}/2) \le 2e^{-c_2d t}$, $\sigma_t\le 1$ and $\|\hth_t-\th_t\|\le 2\l1u$, according to Proposition~\ref{pro:eigmin}.

The next term on the right-hand side of~\eqref{eq:PredictionB2} is bounded in the following proposition.
\begin{propo}\label{pro:lambda_reciprocal}
Using rule~\eqref{lambdat} for $\lambda_t$, we have
\begin{align}
\E\left[\frac{1}{t\lambda_t}\right]\le \frac{6}{\ell_M}\left(\tilde{c} d^2\log T+\frac{4d}{C_{\min}} \log T\right)\,,\label{term2}
\end{align}
where $\tilde{c} = \max(c_1,1/c_2)$ and constants $c_1$ and $c_2$ are defined in Proposition \ref{pro:eigmin} .
\end{propo}
Finally, for the last term, we note that $Q_t$ is rank deficient for $t\le d$ and hence $\sigma_t = 0$, for $1\le t\le d$. Further, the minimum eigenvalue of a matrix is a concave function over PSD matrices. By Jensen inequality, we have 
\begin{align}
\E (\lambda_t) &= \frac{\ell_M}{6t} (1+\sum_{\ell=1}^t \E(\sigma_\ell)) = \frac{\ell_M}{6t} \Big(1+\sum_{\ell=d+1}^t \E(\sigma_\ell)\Big)\nonumber \\ 
&\le \frac{\ell_M}{6t} \Big(1+\sum_{\ell=d+1}^t \sigma_{\min}\Big)  \le \frac{\ell_M}{6t} \Big(1+\frac{t-d}{d}\Big) = \frac{\ell_M}{6d}\,.
\end{align}
In the last inequality, we used the fact that $\Tr(\Sigma) = \E(\|x_t\|^2) = 1$, and thus $\sigma_{\min}\le 1/d$.
Hence,
\begin{align}
\sum_{t=1}^T \E[t\lambda_t] \delta_t \le \frac{\ell_M}{6d} \sum_{t=1}^T  t\delta_t\,, \label{term3}
\end{align}

Using Equations~\eqref{term1}, \eqref{term2}, \eqref{term3} to bound the right-hand side of~\eqref{eq:PredictionB2}, we get
\begin{align}
\E\left[\sum_{t=1}^T \<x_t,\hth_t-\th_t\>^2 \cdot\ind_\cG \right] \le&
\left[\frac{64}{\ell_M^2} + \frac{12u_M^2}{\ell_M^2} \left(\tilde{c}+ \frac{4}{C_{\min} d} \right) \right]\cdot d^2 \log T\nonumber\\
& +\frac{48\l1u^2}{\ell_M^2} + \frac{4\l1u^2 }{c_2 d} +\frac{2\l1u}{d} \sum_{t=1}^T t\delta_t + \frac{\sigma_{\min}}{2}\sum_{t=1}^T  \E(\|\th_t-\hth_t\|^2)\,. \label{term4}
\end{align}

Combining bounds~\eqref{eq:PredictionB1},\eqref{eq:term0} and \eqref{term3}, we obtain
\begin{align*}
\frac{\sigma_{\min}}{2}\sum_{t=1}^T  \E(\|\th_t-\hth_t\|^2)  \le&
\left[\frac{64}{\ell_M^2} + \frac{12u_M^2}{\ell_M^2} \left(\tilde{c}+ \frac{4}{C_{\min} d} \right) \right]\cdot d^2 \log T\\
& + 4\l1u^2 \left(\frac{1}{T} +\frac{12}{\ell_M^2} + \frac{1}{c_2 d} \right) + \frac{2\l1u}{d} \sum_{t=1}^T t\delta_t \,.
\end{align*}
The result follows by recalling that $\sigma_{\min}\ge C_{\min}/d$ as stated by Assumption~\ref{SMF}.

%
%*******************************************
%
\section*{Acknowledgements}

The author was partially
supported by a Google Faculty Research Award.

\newpage
\appendix
\section{Proof of Lemma~\ref{lem:M}}\label{app:M}
We first state some properties of the the virtual valuation function $\varphi$ and the price function $g$, given by Equation~\eqref{eq:g}.
\begin{propo}\label{propo:M}
If $1-F$ is log-concave, then the virtual valuation function $\varphi$ is strictly monotone increasing and the price function $g$ satisfies $0<g'(v)<1$, for
all values of $v\in \reals$. 
\end{propo}
We refer to~\cite{JavanmardNazerzadeh} (Lemmas 1 and 2 in Appendix A therein) for a proof of Proposition~\ref{propo:M}.

For $\th\in \Theta$ we have $\|\th\|\le \l1u$ and hence $|\<x_t,\th\>|\le \|x_t\| \|\th\| \le \l1u$ for all $t$. Applying Proposition~\ref{propo:M} (1-Lipschitz property of $g$),
\[
p_t = g(\<x_t,\th_t\>) \le g(0) + |\<x_t,\th_t\>| \le \varphi^{-1}(0) + \l1u\,.
\]
Therefore,
\begin{align}\label{eq:utB}
|u_t(\th)|\le |p_t| + |\<x_t,\th\>|\le \varphi^{-1}(0) + 2\l1u\,.
\end{align}

\section{Proof of Lemma~\ref{lem:DUB}}\label{app:DUB}

We note that the update rule~\eqref{eq:ML} can be recast as $\hth_{t+1} = \arg\min_{\th\in \Theta} \cC_t(\th)$, where 
$$\cC_t(\th) = \eta_t\<\nabla \ell_t(\hth_t),\th\>+\frac{1}{2}\|\th-\hth_t\|^2\,.$$
By convexity of $\cC_t$ and optimality of $\hth_{t+1}$, we have $\<\th-\hth_{t+1},\nabla \cC_t(\hth_{t+1})\>\ge 0$ for all $\th\in \Theta$. Setting $\th=\th_t$,
\begin{align}
\<\th_t-\hth_{t+1}, \eta_t \nabla \ell_t(\hth_t) + \hth_{t+1}-\hth_{t}\> \ge 0\,.\label{eq:E1}
\end{align}

Expanding $\ell_t(\th)$ around $\hth_t$, we have
\begin{align}\label{eq:E2}
\ell_t(\hth_t)-\ell(\th_t) = \<\nabla\ell_t(\hth_t), \hth_t-\th_t\> -\frac{1}{2}\<\th_t-\hth_t, \nabla^2\ell_t(\tth) (\th_t-\hth_t)\> \,,
\end{align}
for some $\tilth$ on the line segment between $\tth_t$ and $\hth_t$. 
%Throughout, we use notation $\<u,v\>$ to refer to the inner product of two vectors
%$u$ and $v$. 
Recalling~\eqref{eq:lt}, the gradient and the hessian of $\ell_t$ read as
\begin{align}\label{eq:nabla-nabla2}
\nabla \ell_t(\th) = \mu_t(\th) x_t\,, \quad \nabla^2 \ell_t(\th) = \eta_t(\th) x_tx_t^\sT\,,
\end{align}
with,
\begin{eqnarray}
\mu_t(\th) &=& -\frac{{f}(u_t(\th))}{F(u_t(\th))}\ind(y_t = -1) + \frac{{f}(u_t(\th))}{1-F(u_t(\th))} \ind(y_t = +1) \nonumber\\
&=& -\dx \log F(u_t(\th)) \ind(y_t = -1) - \dx \log(1-F(u_t(\th))) \ind(y_t = +1)\label{eq:mu}
\end{eqnarray}
\begin{eqnarray}
\eta_t(\th)  &=&  \bigg(\frac{f(u_t(\th))^2}{F(u_t(\th))^2} - \frac{{f'}(u_t(\th))}{F(u_t(\th))} \bigg)\ind(y_t = -1) +\bigg(\frac{f(u_t(\th))^2}{(1-F(u_t(\th)))^2}+ \frac{{f'}(u_t(\th))}{1-F(u_t(\th))} \bigg)\ind(y_t = +1)\nonumber\\
&=& -\ddx \log F(u_t(\th)) \ind(y_t = -1) - \ddx \log (1-F(u_t(\th))) \ind(y_t = +1)\,. \label{eq:eta}
\end{eqnarray}
Here, $u_t(\th) = p_t-\<x_t,\th\>$, and $\dx \log F(x)$ and $\ddx \log F(x)$ represent first and second derivative w.r.t $x$, respectively.
In addition, using Equation~\eqref{eq:utB}
\begin{align}\label{eq:utB}
|u_t(\th)| \le \varphi^{-1}(0) + 2\l1u = M\,, \quad \forall \th\in \Theta\,. 
\end{align}
Hence, invoking the definition of $\ell_M$, as per Equation~\eqref{eq:lM}, we get that $\eta_t(\th) \ge \ell_M$ and hence $\nabla^2\ell_t(\tilth) \succeq \ell_M x_t x_t^\sT$.

Continuing from Equation~\eqref{eq:E2}, we get
\begin{align}\label{eq:E3}
\ell_t(\hth_t)-\ell(\th_t) &\le \<\nabla\ell_t(\hth_t), \hth_t-\th_t\> -\frac{\ell_M}{2}\<x_t,\th_t-\hth_t\>^2 \nonumber\\
&= \<\nabla\ell_t(\hth_t), \hth_{t+1}-\th_t\> + \<\nabla\ell_t(\hth_t), \hth_t-\hth_{t+1}\> -\frac{\ell_M}{2}\<x_t,\th_t-\hth_t\>^2\nonumber\\
&\le \frac{1}{\eta_t} \<\th_t-\hth_{t+1},\hth_{t+1}-\hth_t\>+ \<\nabla\ell_t(\hth_t), \hth_t-\hth_{t+1}\> -\frac{\ell_M}{2}\<x_t,\th_t-\hth_t\>^2\nonumber\\
&= \frac{1}{2\eta_t} \Big\{\|\th_t-\hth_t\|^2-\|\th_t-\hth_{t+1}\|^2-\|\hth_{t+1} - \hth_t\|^2 \Big\} \nonumber\\
&\quad \,\,+  \<\nabla\ell_t(\hth_t), \hth_t-\hth_{t+1}\> -\frac{\ell_M}{2}\<x_t,\th_t-\hth_t\>^2\nonumber\\
&=\frac{1}{2\eta_t}\Big\{\|\th_t-\hth_t\|^2-\|\th_{t+1}-\hth_{t+1}\|^2\Big\} + \frac{1}{2\eta_t}\Big\{\|\th_{t+1}-\hth_{t+1}\|^2 - \|\th_t-\hth_{t+1}\|^2 \Big\} \nonumber\\
&\quad\,\, -\frac{1}{2\eta_t} \|\hth_{t+1} - \hth_t\|^2 +  \<\nabla\ell_t(\hth_t), \hth_t-\hth_{t+1}\> -\frac{\ell_M}{2}\<x_t,\th_t-\hth_t\>^2
\end{align} 
We next note that the second term above can be bounded as
\begin{align}\label{eq:E4}
\frac{1}{2\eta_t} \Big\{\|\th_{t+1}-\hth_{t+1}\|^2 - \|\th_t-\hth_{t+1}\|^2 \Big\}  = \frac{1}{\eta_t}\<\th_{t+1}-\hth_{t+1}, \th_{t+1}-\th_t\> \le \frac{2}{\eta_t} \l1u \delta_t\,,
\end{align}
because $\th_{t+1}, \hth_{t+1}\in \Theta$ and hence $\|\th_{t+1}-\hth_{t+1}\|\le 2\l1u$ by triangle inequality. 

Further,
\begin{align}\label{eq:E5}
\<\nabla\ell_t(\hth_t), \hth_t-\hth_{t+1}\> &\le  \frac{1}{2\eta_t} \|\hth_{t+1} - \hth_t\|^2 + \frac{\eta_t}{2}\|\nabla \ell_t(\hth_t)\|^2 \nonumber\\
&\le  \frac{1}{2\eta_t} \|\hth_{t+1} - \hth_t\|^2 + \frac{\eta_t}{2} |\mu(\hth_t)|^2 \|x_t\|^2 \le  \frac{1}{2\eta_t} \|\hth_{t+1} - \hth_t\|^2  + \frac{\eta_t}{2} u_M^2\,,
\end{align}
where we used the inequality $2ab\le a^2+b^2$ and the characterization of gradient~\eqref{eq:nabla-nabla2}. Note that by~\eqref{eq:utB}, $|u_t(\hth)|\le M$ and by definition~\eqref{eq:uM}, $|\mu_t(\hth_t)|\le u_M$. 
Plugging in bounds from~\eqref{eq:E4} and~\eqref{eq:E5} in Equation~\eqref{eq:E3}, we arrive at
\begin{align}
\ell_t(\hth_t)-\ell(\th_t) \le \frac{1}{2\eta_t}\Big\{\|\th_t-\hth_t\|^2-\|\th_{t+1}-\hth_{t+1}\|^2\Big\} + \frac{2}{\eta_t} \l1u \delta_t +  \frac{\eta_t}{2}u_M^2  -\frac{\ell_M}{2}\<x_t,\th_t-\hth_t\>^2
\end{align}
We use the shorthand $D_t = \frac{1}{2} \|\th_t-\hth_t\|^2$. The result follows by summing the above bound over time:
\begin{align*}
\sum_{t=1}^T \ell_t(\hth_t) - \sum_{t=1}^T \ell_t(\th_t) =& \sum_{t=1}^T \Big(\frac{D_t}{\eta_t} - \frac{D_{t+1}}{\eta_{t+1}} \Big) + \sum_{t=1}^T D_{t+1}\Big(\frac{1}{\eta_{t+1}}-\frac{1}{\eta_t} \Big) \\
&+\frac{u_M^2}{2} \sum_{t=1}^T \eta_t + 2\l1u \sum_{t=1}^T \frac{\delta_t }{\eta_t} -\frac{\ell_M}{2}\sum_{t=1}^T\<x_t,\th_t-\hth_t\>^2\,.
\end{align*}
The proof is concluded because $D_1\le 2\l1u^2$ as $\hth_1,\th_1\in \Theta$; therefore
\begin{align*}
 \sum_{t=1}^T \Big(\frac{D_t}{\eta_t} - \frac{D_{t+1}}{\eta_{t+1}} \Big)  = \frac{D_1}{\eta_1}-\frac{D_{T+1}}{\eta_{T+1}} \le \frac{D_1}{\eta_1} \le \frac{2\l1u^2}{\eta_1}\,.
\end{align*}
%

%====================================
\section{Proof of Lemma~\ref{lem:DLB}}\label{app:DLB}
By convexity of $\ell_t(\th)$, we have
\begin{align}\label{Eq6}
\ell_t(\th_t) - \ell_t(\hth_t) \le \<\nabla \ell_t(\th_t),\hth_t-\th_t\> = \mu_t(\th_t) \<x_t,\th_t-\hth_t\>\,.
\end{align}
We denote $D_t = \mu_t(\th_t) \<x_t,\th_t-\hth_t\>$ and let $\cF_t$ be the $\sigma$-algebra generated by $\{z_t\}_{t=1}^T$. Since $\hth_t$ is $\cF_{t-1}$ measurable, we have
\begin{align}
\E(D_t|\cF_{t-1}) = \E(\mu_t(\th_t)|\cF_{t-1}) \<x_t,\th_t-\hth_t\> = 0\,,
\end{align}
where $\E(\mu_t(\th_t)|\cF_{t-1}) = 0$ follows readily from Equation~\eqref{eq:mu}. Therefore, $D(T) \equiv \sum_{t=1}^T D_t$ is  a martingale adapted to the filtration $\cF_{t}$.

We next bound $\E[e^{\lambda D_t}|\cF_{t-1}]$ for any $\lambda\in \reals$. Conditional on $\cF_{t-1}$, we have $|D_t|\le \beta_t$, with $\beta_t \equiv u_M|\<x_t,\th_t-\hth_t\>|$. Since $e^{\lambda z}$ is convex,
\begin{align}
\E[e^{\lambda D_t}|\cF_{t-1}] &\le \E\bigg[\frac{\beta_t-D_t}{2\beta_t}  e^{-\lambda \beta_t}  +\frac{\beta_t+D_t}{2\beta_t} e^{\lambda \beta_t}\bigg| \cF_{t-1} \bigg]\nonumber\\
&=\E\bigg[\frac{e^{-\lambda \beta_t}+e^{\lambda \beta_t}}{2} \bigg] + \E[D_t|\cF_{t-1}] \bigg(\frac{e^{-\lambda \beta_t}+e^{\lambda \beta_t}}{2\beta_t} \bigg) = \cosh(\lambda \beta_t) \le e^{\lambda^2 \beta_t^2/2}\,.
\end{align}
We are now ready to apply the following Bernstein-type concentration bound for martingale difference sequences, whose proof is given in Appendix~\ref{app:MD} for the reader's convenience. 
\begin{propo}\label{propo:MD}
Consider a martingale difference sequence $D_t$ adapted to a filtration $\cF_t$, such that for any $\lambda\ge 0$, $\E[e^{\lambda D_t}|\cF_{t-1}] \le e^{\lambda^2\sigma_t^2/2}$\,.
Then, for $D(T) = \sum_{t=1}^T D_t$, the following holds true:
\begin{align}
\prob(D(T)\ge \xi) \le e^{-{\xi^2}/({2\sum_{t=1}^T \sigma_t^2})}\,.
\end{align}
\end{propo}
Combining Equation~\eqref{Eq6} and the result of Proposition~\ref{propo:MD} we obtain 
\begin{align}
\prob\bigg(\sum_{t=1}^T \ell_t(\hth_t) - \sum_{t=1}^T\ell_t(\th_t) \le- 2\sqrt{\log T} \Big\{\sum_{t=1}^T \<x_t,\th_t-\hth_t\>^2\Big\}^{1/2} \bigg) \le \frac{1}{T^2}\,.
\end{align}
The result follows.  

%============================
\section{Proof of Proposition~\ref{propo:MD}}\label{app:MD}
We follow the standard approach of controlling the moment generating function  of $D(T)$.Conditioning on $\cF_{t-1}$ and applying iterated expectation yields
\begin{align}
\E[e^{\lambda D(T)} ] = \E\bigg[e^{\lambda \sum_{t=1}^{T-1} D_t} \cdot \E[e^{\lambda D_T}|\cF_{T-1}] \bigg] \le \E\bigg[e^{\lambda \sum_{t=1}^{T-1} D_t} \bigg]  e^{\lambda^2 \sigma_T^2/2}\,.
\end{align}
Iterating this procedure gives the bound $\E[e^{\lambda \sum_{t=1}^T D_t}]\le e^{\lambda^2 \sum_{t=1}^T \sigma_t^2/2}$, for all $\lambda \ge 0$.

Now by applying the exponential Markov inequality we get
\begin{align}
\prob(D(T)\ge \xi) = \prob(e^{\lambda D(T)} \ge e^{\lambda \xi})\le e^{-\lambda \xi} \E[e^{\lambda \sum_{t=1}^T D_t}] \le e^{-\lambda \xi} e^{\lambda^2 (\sum_{t=1}^T \sigma_t^2)/2}\,.
\end{align}
Choosing $\lambda = \xi/(\sum_{t=1}^T \sigma_t^2)$ gives the desired result.

%=============================================
\section{Proof of Proposition~\ref{pro:eigmin}}
We prove the result in a more general case, namely when the features are independent random vectors with bounded subgaussian norms. 
\begin{definition}
For a random variable $z$, its subgaussian norm, denoted by $\|z\|_{\psi_2}$ is defined as
\begin{align}
\|z\|_{\psi_2} = \sup_{p\ge 1} \;p^{-1/2} (\E|z|^p)^{1/p}\,.
\end{align}
Further, for a random vector $z$ its subgaussian norm is defined as
\begin{align}
\|z\|_{\psi_2} = \sup_{\|u\|\ge 1} \; \|\<z,u\>\|_{\psi_2}\,.
\end{align}
\end{definition} 
We next recall the following result from~\cite{vershynin} about random matrices with independent rows.

\begin{propo}\label{versh}
Suppose $x_\ell\in \reals^d$ are independent random vectors generated from a distribution with covariance $\Sigma$
and their subgaussian norms are bounded by $K$. Further, let $Q_t = (1/t) \sum_{\ell=1}^t x_\ell x_\ell^\sT$. 
Then for every $s\ge 0$, the following inequality holds with probability at least $1-2\exp(-cs^2)$:
\begin{align}
\Big\|Q_t-\Sigma \Big\| \le \max(\delta,\delta^2) \quad \quad \quad \text{ where }\delta = C\sqrt{\frac{d}{t}} + \frac{s}{\sqrt{t}}\,. 
\end{align}
Here $C$ and $c>0$ are constants that depend solely on $K$.
\end{propo}

We next show that the feature vectors in our problem have bounded subgaussian norm. Given that $\|x_\ell\|\le 1$, 
for $\|u\|\le 1$, we have
\begin{align*}
\|\<x_\ell,u\>\|_{\psi_2} =\sup_{p\ge 1}\; p^{-1/2} (\E|\<x_\ell,u\>|^p)^{1/p}\le \sup_{p\ge 1}\; p^{-1/2} (\E[\|x_\ell\|\|u\|]^p)^{1/p}\le 1\,.
\end{align*}

Applying Proposition~\eqref{versh} with $K=1$, there exist constants $c_1,c_2$ (depending on $C_{\min}$), such that for $t\ge c_1d^2$, we have
\begin{align}\label{pert}
\|Q_t -\Sigma\| \le \frac{1}{2d} C_{\min} \le \frac{1}{2} \sigma_{\min}\,,
\end{align}
with probability at least $1-2e^{-c_2 t/d}$. Weyl's inequality then implies that $|\sig_t-\sigma_{\min}|\le \sigma_{\min}/2$.   
%By union bounding, we obtain that
%%
%\begin{align*}
%\prob\Big(\forall t\ge Cd:\; \frac{1}{2}\sigma_{\min} \le \sig_t \le \frac{3}{2}\sigma_{\min} \Big)\ge 1- 2\sum_{t=c_1d}^\infty e^{-c_2t} \ge 1-\frac{2}{c_2}e^{-c_1c_2d}\,.
%\end{align*}
%%
%This completes the proof for subgaussian random vectors. 

Also note that for $t\ge 1$,
\[
\sigma_t\le \|Q_t\| \le \frac{1}{t}\sum_{\ell=1}^t \|x_\ell x_\ell^\sT\|  = \frac{1}{t}\sum_{\ell=1}^t \|x_\ell\|^2  = 1\,.
\]
The proof is complete.

%======================================
\section{Proof of Lemma~\ref{pro:lambda_reciprocal}} \label{app:lambda_reciprocal}

The way we set $\lambda_t$ (see Equation~\eqref{lambdat}), we have
\[\frac{1}{t\lambda_t} = \left(\frac{6}{\ell_M}\right) \frac{1}{1+\sig_1+\sig_2+\dotsc+\sig_t} \]

Clearly, for $t\ge 1$, $1/(t\lambda_t)\le 6/\ell_M$. Let $t_0 = \tilde{c} d^2 \log T$, with $\tilde{c}= \max(c_1,1/c_2)$. For $T\ge t_0$, define the event $\cE_T$ as follows
\begin{align}
\cE_T = \{\sigma_t \ge \sigma_{\min}/2, \, \text{for } t_0\le t\le T\}\,.
\end{align}
By applying Proposition~\ref{pro:eigmin} and union bounding over $t$, we get
\begin{align}
\prob(\cE_T) \ge 1 - \sum_{t = t_0}^T 2e^{-c_2 t/d} \ge 1 - \frac{2d}{c_2} e^{-c_2t_0/d} 
\end{align} 
Therefore,  
\begin{align}
\sum_{t=t_0}^T \E\left[\frac{1}{t\lambda}\right] &\le \E\left[\left(\sum_{t=t_0}^T\frac{1}{t\lambda}\right) \ind(\cE_T)\right] + \frac{6T}{\ell_M}  \prob(\cE_T^c)\nonumber\\
& = \frac{6}{\ell_M}\E\left[\left(\sum_{t=t_0}^T\frac{1}{1+\sigma_1+\dotsc+\sigma_t}\right)\cdot \ind(\cE_T)\right] + \frac{6T}{\ell_M} \prob(\cE_T^c)\nonumber\\
&\le  \frac{6}{\ell_M}\left(\sum_{t=1}^T \frac{1}{1+\frac{t}{2}\sigma_{\min}} +\frac{2d}{c_2} T^{1-c_2\tilde{c}d}\right)\nonumber\\
&\le \frac{12}{\ell_M} \left(\frac{1}{\sigma_{\min}} \log T + \frac{d}{c_2} T^{1-d} \right) \le  \frac{24d}{\ell_M C_{\min}} \log T\,.
\end{align}
For $t\ge 1$, we use the bound $1/(t\lambda_t)\le 6/\ell_M$. Hence,
\begin{align}
\sum_{t=1}^T \E\left[\frac{1}{t\lambda}\right] \le  \frac{6}{\ell_M}\left(t_0+\frac{4d}{C_{\min}} \log T\right) \le \frac{6}{\ell_M}\left(\tilde{c} d^2\log T+\frac{4d}{C_{\min}} \log T\right)
\end{align}

The proof is complete.
\medskip
%\section*{Acknowledgment}
%\aj{Anything here?}
%\paragraph{\bf Acknowledgments}
%We are thankful to Arnoud den Boer, Mohsen Bayati, and Paat Rusmevichientong for their suggestions that improved this work.

\bibliographystyle{amsalpha}
\bibliography{dynamicpricing}
\end{document}

%% file: notation.tex
\def\tp{\tilde{p}}

\def\bx{\boldsymbol{x}}
\def\KL{D_{\rm KL}}
\def\cV{\mathcal{V}}
\def\bth{\boldsymbol{\theta}}

\def\cF{{\cal F}}

\def\cC{{\cal C}}
\def\cG{{\cal G}}
\def\cE{{\cal E}}

\def\tp{{\tilde{p}}}

\def\cG{\mathcal{G}}

\def\dis{{\sf d}}

\def\naturals{{\mathbb N}}

\def\reals{{\mathbb R}}

\def\prob{{\mathbb P}}
\def\E{{\mathbb E}}

\def\tZ{\tilde{Z}}

\def\L0{{L_i}}

\def\de{{\rm d}}
\def\<{\langle}
\def\>{\rangle}

\def\dx{\frac{\de}{\de x}}
\def\ddx{\frac{\de^2}{\de x^2}}

\def\hth{\widehat{\theta}}

\def\F{{\sf F}}
\def\ind{{\mathbb I}}

\def \Tr{{\rm Trace}}
\def\F{{\sf F}}
\def\normal{{\sf N}}

\def\sT{{\sf T}}

\def\id{{\rm I}}

\def\v*{v_i}
\def\T*{T_i}

\def\u*{u_i}
\def\F*{F_i}

\definecolor{olivegreen}{rgb}{0,0.6,0.4}

\def\tilth{\tilde{\theta}}

\def\th{{\theta}}
\def\hth{{\widehat{\theta}}}
\def\tth{{\theta^*}}

\def\cX{\mathcal{X}}

\def\Reg{{\sf Regret}}
\def\popt{p^*}

\def\l1u{W}

\def\tth{\widetilde{\theta}}
\def\sig{\sigma}

\def\ctF{\mathcal{\tilde{F}}}

\newcommand{\ajcomment}[1]{}

\makeatletter
\newcommand{\labitem}[2]{%
\def\@itemlabel{\text{#1}}
\item
\def\@currentlabel{#1}\label{#2}}
\makeatother

\DeclareMathAlphabet{\mathpzc}{OT1}{pzc}{m}{it}